\documentclass[aps,pre,showpacs,floatfix]{revtex4-1}
\usepackage{graphicx}
\usepackage{dcolumn}
\usepackage{bm}
\usepackage{epsf}
\usepackage{subfigure}
\usepackage{epstopdf}
\usepackage{amsmath}
\usepackage{amssymb}
\usepackage[cspex,bbgreekl]{mathbbol}
\usepackage{color}
\usepackage{stackrel}
\newcommand{\bea}{\begin{eqnarray}}
\newcommand{\eea}{\end{eqnarray}}

\newcommand{\trace}{{\rm Tr}}

\newcommand{\be}{\begin{equation}}
\newcommand{\ee}{\end{equation}}
\newcommand{\ba}{\begin{eqnarray}}
\newcommand{\ea}{\end{eqnarray}}
\newcommand{\s}{\boldsymbol \sigma}
\newcommand{\M}{{\bf M}}

\newcommand{\n}{{\bf n}}

\newcommand{\m}{{\bf m}}
\newcommand{\B}{{\bf B}}
\newcommand{\sr}{{\bf r}}

\graphicspath{{Code-Python/}}
\graphicspath{{figure/}} 

\begin{document}

\title{Fluctuation relations for equilibrium states \\
with broken discrete or continuous symmetries}

\author{D. Lacoste$^1$}
\author{P. Gaspard$^2$}

\affiliation{$^1$ Laboratoire de Physico-Chimie Th\'eorique - UMR CNRS Gulliver 7083,\\ PSL Research University, ESPCI, 10 rue Vauquelin, F-75231 Paris, France\\
$^2$ Center for Nonlinear Phenomena and Complex Systems, \\Universit\'e Libre de Bruxelles, Code Postal 231, Campus Plaine, B-1050 Brussels, Belgium}

\date{\today}

\begin{abstract}
Isometric fluctuation relations are deduced for the fluctuations of the order parameter in equilibrium systems of condensed-matter physics with
 broken discrete or continuous symmetries.  These relations are similar to their analogues obtained for non-equilibrium systems where the broken 
symmetry is time reversal.  At equilibrium, these relations show that the ratio of the probabilities of opposite fluctuations goes exponentially 
with the symmetry-breaking external field and the magnitude of the fluctuations.
These relations are applied to the Curie-Weiss, Heisenberg, and $XY$~models of magnetism where the continuous rotational symmetry is broken, 
as well as to the $q$-state Potts model and the $p$-state clock model where discrete symmetries are broken. 
Broken symmetries are also considered in the anisotropic Curie-Weiss model.  
For infinite systems, the results are calculated using large-deviation theory. The relations are also applied to mean-field models of nematic
 liquid crystals where the order parameter is tensorial.  Moreover, their extension to quantum systems is also deduced.
\end{abstract}


\maketitle

\section{Introduction}

At macroscopic scales, the second law of thermodynamics characterizes the breaking of the 
time-reversal symmetry due to energy dissipation in non-equilibrium systems.
At microscopic scales, the time-reversal symmetry still holds, and this has various important 
consequences for fluctuations. One of them is the existence of symmetry relations called fluctuation relations, 
which constrain the probability distributions of thermodynamic quantities  
arbitrarily far from equilibrium. The discovery of fluctuation relations represents
a major progress in our understanding of the second law of thermodynamics and has also accompanied 
many advances in the observation and manipulation of experimental non-equilibrium systems 
\cite{ECM93,GC95,K98,C99,LS99,M99,AG06JSM,AG07,EHM09,J11,S12}.

All these studies have put a strong emphasis on 
dissipative systems with broken time-reversal symmetry, despite the fact that 
many other forms of symmetry breaking are known in nature. In fact, the concept 
is so central that it enters practically all areas of science \cite{N09,GSW62,A72,A84,F75,CL95,BE07,H14,PN67,PLGH69,CH93,PJ15}.  
It seems therefore rather important to explore the general connection between fluctuation theorems and 
symmetry breaking, when considering symmetries not related to time. 
Conveniently, this can be done with equilibrium systems,  
 which are much better understood than non-equilibrium systems. 
By studying equilibrium systems from the viewpoint of non-equilibrium systems, 
the objective is to gain further insights on the thermodynamics of non-equilibrium systems. 
At the same time, this will contribute to clarify the deep connections 
which exist between fluctuations and 
symmetries, while suggesting new ideas of methods for extracting relevant information 
from the fluctuations of equilibrium systems.

A symmetry may be broken spontaneously if the ground state has a lower symmetry than the Hamiltonian, 
i.e., if a perturbation $H_1$ is added to some Hamiltonian $H_0$ where $H_1$ is less symmetric than $H_0$.  In such circumstances, we may wonder if the fluctuations 
of the order parameter leave a footprint of the symmetry that is broken.
A related question was raised by Goldenfeld in his famous lectures given in the sixties \cite{GF92} in an attempt to understand spontaneous symmetry breaking at the level of probability distributions.
Considering the Ising model in the presence of a magnetic field $B$, he observed that the 
ratio of the probabilities to be in the two symmetry broken states of opposite magnetizations $\pm M$ obeys the relation
\be
\label{Goldenfeld}
\frac{P_B(M)}{P_B(-M)}={\rm e}^{2 \beta B M},
\ee
which follows immediately from the presence of a coupling term linear in the magnetic field in the Hamiltonian together with Boltzmann's distribution.
This relation has interesting implications for spontaneous symmetry breaking (SSB) on which we shall come back later in this paper.
The similarity of Eq.~(\ref{Goldenfeld}) with fluctuation theorems discovered for non-equilibrium systems has only been noticed
 recently in Refs.~\cite{K10,G12PS,G12JSM}.

By considering the symmetry under both time reversal and spatial rotations in non-equilibrium fluids, Hurtado {\it et al.}
 uncovered in 2011 a remarkable extension of the fluctuation relation for vectorial currents, which they dubbed {\it isometric fluctuation relations} 
\cite{HPPG11,HPPG14,HPPG15}.
These results hint at the possibility that all the fundamental symmetries continue to 
manifest themselves in the fluctuations, even if these symmetries are broken by external constraints.
This concerns not only systems driven away from equilibrium, but also equilibrium systems.

In our recent work on this topic \cite{LG14}, we have combined these results and generalized Eq.~(\ref{Goldenfeld}) using the language of group theory to describe the symmetry of the Hamiltonian. 
We have found that for equilibrium systems, whenever a symmetry is broken by an external field, the probability distribution of the fluctuations obeys an isometric fluctuation relation of this type.

In this longer paper, we provide much more details on this topic, and we illustrate the relations on a larger number of models of statistical physics.  
In Section~\ref{IFR}, we present the general proof of isometric fluctuation relations for finite systems.  We also discuss many direct implications of the relation 
and we extend the relations to quantum systems. In Section~\ref{IFRinf}, we present the form of the relation 
in the thermodynamic limit, where it is related to the notion of large-deviation function \cite{E85,E95,D07,T09}. We also 
show the implications of the relation for spontaneous symmetry breaking.
In Section~\ref{Magnetic}, we present results for magnetic systems, such as the Curie-Weiss model, the one-dimensional Heisenberg chain, and the $XY$ model.  Then Section~\ref{Anisotropic} deals with anisotropic systems described by subgroups of continuous groups, while Section~\ref{Nematic} covers the case of nematic liquid crystals.  Conclusions are drawn in Section~\ref{Concl}.

\section{Isometric fluctuation relations in finite systems}
\label{IFR}

\subsection{Derivation of a general identity in the canonical ensemble}
\label{IFR-begin}

Let us consider a system composed of $N$ classical spins $\s=\{ \pmb{\sigma}_i \}_{i=1}^N$ taking 
discrete or continuous values such that $\pmb{\sigma}_i\in{\mathbb R}^d$ and 
$\Vert\pmb{\sigma}_i\Vert=1$.  The Hamiltonian of the system is assumed to be of the form
\be
H_N(\s;\B)=H_N(\s;{\bf 0})- \B \cdot \M_N(\s)
\label{Hamilt}
\ee 
where $\B$ is the external magnetic field and the order parameter is the magnetization 
\be
\M_N(\s)=\sum_{i=1}^N \s_i \, .  
\label{Magnet}
\ee
We suppose that the system is at equilibrium in the Gibbsian canonical distribution at the inverse
 temperature~$\beta$ \cite{P03}
\be
\mu_{\bf B}(\s)= \frac{1}{Z_N(\B)} \, {\rm e}^{- \beta H_N(\s;\B)} ,
\label{canonical}
\ee
where $Z_N(\B)=\sum_{\pmb{\sigma}} {\rm e}^{- \beta H_N(\s;\B)}$ is the classical partition function such that the distribution is normalized to unity: $\sum_{\pmb{\sigma}}\mu_{\bf B}(\s)=1$.

Let us also introduce an observable function $A(\s)$ of the spin variables $\s\in{\mathbb R}^{Nd}$.
The function $A(\s)$ is assumed to be scalar for simplicity.  For this observable function, we establish a general identity for the system in the presence and the absence of the external field $\B$.  Denoting by $\langle \cdot \rangle_\B$ the statistical average over the probability distribution $\mu_{\bf B}(\s)$, we find that
\ba
\langle A(\s)\, {\rm e}^{-\beta \B \cdot \M_N(\s)} \rangle_\B &=& \sum_{\s} \mu_{\bf B}(\s)\, A(\s)\, {\rm e}^{-\beta \B \cdot \M_N(\s)}\, , \nonumber\\
&=& \frac{1}{Z_N({\bf B})} \sum_{\s} {\rm e}^{-\beta H_N(\s;{\bf 0})} A(\s)\, ,  
\nonumber\\
&=&\frac{Z_N({\bf 0})}{Z_N({\bf B})}\; \langle A(\s) \rangle_{\bf 0}\, .
\label{Extension-gen-observable}
\ea
This general identity can be rewritten in the form
\be
\langle A(\s)\, {\rm e}^{-\beta \B \cdot \M_N(\s)} \rangle_\B =  {\rm e}^{-\beta \Delta F}  \langle A(\s) \rangle_{\bf 0}
\ee
in terms of the difference of free energy $\Delta F=F_N({\bf 0})-F_N(\B)$ between the states with $\B={\bf 0}$ and the state with a non-zero magnetic field $\B$, because the free energy is related to the partition function by $Z_N({\B})={\rm e}^{-\beta F_N(\B)}$.

In the particular case where $A(\s)=1$, one gets
\be
\label{Jarzynski}
\langle {\rm e}^{-\beta \B \cdot \M_N(\s)} \rangle_\B ={\rm e}^{-\beta \Delta F}\, ,
\ee
which makes apparent the similarities between our relation~(\ref{Extension-gen-observable}) and the Jarzynski relation \cite{J11}. Note that, in this analogy, the Jarzynski work is replaced by the part of the Hamiltonian that is due to the symmetry breaking, namely $\B \cdot \M_N(\s)$, and $\B$ is the control parameter.

\subsection{Fluctuations of the order parameter}

Let us also define the probability density $P_\B(\M)$ that the magnetization takes the value $\M=\M_N(\s)$ as
\be
P_\B(\M) \equiv \langle \delta\left[\M-\M_N(\s)\right]\rangle_\B
\ee
where $\delta(\cdot)$ denotes the Dirac delta distribution and $\langle \cdot\rangle_\B$ the statistical average over Gibbs' canonical measure (\ref{canonical}). 
This probability density is a function of the vectorial magnetization $\M\in{\mathbb R}^d$ and it 
is normalized according to
\be
\int d\M \, P_\B(\M) = 1 \, .
\label{normalization}
\ee

If we take $A(\s)=\delta\left[\M-\M_N(\s)\right]$ in the general identity~(\ref{Extension-gen-observable}), we obtain an identity between the distribution of the order parameter in the field, $P_{\bf B}({\bf M})=\langle A(\s)\rangle_{\B}$, and the same distribution in the absence of the field, $P_{\bf 0}({\bf M})=\langle A(\s)\rangle_{\bf 0}$:
\be
P_\B(\M) = \frac{Z_N({\bf 0})}{Z_N({\bf B})}\; {\rm e}^{\beta \B \cdot \M} \; P_{\bf 0}({\bf M}),
\label{GI}
\ee
which we have previously deduced in Ref.~\cite{LG14}.  
Expressing the ratio of partition functions in terms of the difference of free energy $\Delta F$, it follows from Eq.~(\ref{GI}) that
\be
\label{Crooks}
\frac{P_{\bf B}(\M)}{P_{\bf 0}(\M)} = {\rm e}^{\beta (\B \cdot \M -\Delta F)}\, ,
\ee
which is the equilibrium analogue of the Crooks fluctuation theorem \cite{C99}.  From this Crooks-like relation, the Jarzynski-like relation of Eq.~(\ref{Jarzynski}) follows directly.  In analogy with the Jarzynski and Crooks relations which allow us to estimate free energies from non-equilibrium fluctuations of the work, 
Eq.~(\ref{Jarzynski}) and Eq.~(\ref{Crooks}) could be used to estimate $\Delta F$ 
from measurements of equilibrium fluctuations of the magnetization $\M$, 
and by extension from measurements of the fluctuations for other relevant order parameter \cite{DL15}.

An important remark is that the previous identities hold even if the Hamiltonian $H_N(\s;{\bf 0})$ has no particular symmetry.

\subsection{Derivation of the isometric fluctuation relations}

Now, the Hamiltonian $H_N(\s;{\bf 0})$ is supposed to be invariant under a symmetry group~$G$ in the absence of external field.
Accordingly, we have that $H_N(\pmb{\sigma}^g;{\bf 0})= H_N(\pmb{\sigma};{\bf 0})$, where $\s^g=\{ {\boldsymbol{\mathsf R}}_g\cdot\pmb{\sigma}_i \}_{i=1}^N$,
 and ${\boldsymbol{\mathsf R}}_g$ is a representation of the element~$g$ of the group~$G$.  This group $G$ may be discrete or continuous, and as we shall see
 later, this distinction is crucial for evaluating the properties of the probability distribution of the fluctuations.  Let us emphasize three important points:
 (i) the group acts on the degrees of freedom of the spins; 
(ii) the symmetry that we consider is global and not
 local, since the group acts on all the spins irrespective of their location in the space in which they are embedded; and (iii) 
the groups that we are interested in should satisfy $\vert\det{\boldsymbol{\mathsf R}}_g\vert=1$.

As a consequence, the probability distribution of the magnetization has this symmetry in the absence of magnetic field since summing over the microstates ${\s}$ or their symmetry transforms ${\s^g}$ are equivalent  for every $g\in G$ so that
\ba
P_{\bf 0}({\bf M})&=& \frac{1}{Z_N({\bf 0})} \sum_{\s^g} {\rm e}^{-\beta H_N(\s^g;{\bf 0})} \delta\left[\M-\M_N(\s^g)\right], \nonumber\\
&=& \frac{1}{Z_N({\bf 0})} \sum_{\s} {\rm e}^{-\beta H_N(\s;{\bf 0})} \delta\left[\M-{\boldsymbol{\mathsf R}}_g\cdot\M_N(\s)\right], \nonumber\\
&=& P_{\bf 0}({\boldsymbol{\mathsf R}}_g^{-1}\cdot{\bf M}),
\label{proof}
\ea
where, in the last step, we have used a change of variables in the sum with a Jacobian equal to one thanks to the property $\vert\det{\boldsymbol{\mathsf R}}_g\vert=1$.
Combining Eqs.~(\ref{GI}) and~(\ref{proof}), 
one obtains the fluctuation relation:
\be
P_{\bf B}({\bf M}) = P_{\bf B}({\bf M'}) \ {\rm e}^{\beta \B\cdot({\bf M}-{\bf M'})}. 
\label{FT1}
\ee
with ${\bf M'}={\boldsymbol{\mathsf R}}_g^{-1}\cdot\M$ for all $g\in G$.  
Since a group contains the inverse $g^{-1}$ of any element $g\in G$, Eq.~(\ref{FT1}) also holds with ${\bf M'}={\boldsymbol{\mathsf R}}_g\cdot\M$.

When ${\boldsymbol{\mathsf R}}_g$ represents a rotation, $\Vert{\bf M}\Vert = \Vert{\bf M'}\Vert$, 
hence the name {\it isometric fluctuation relation} attached to this particular case. 
This relation includes as a particular case the fluctuation relation derived in Ref.~\cite{G12PS,G12JSM} when ${\bf M'}=-\M$
corresponding to the ${\mathbb Z}_2$ group. 
However, as illustrated in the next sections of this paper, other representations ${\boldsymbol{\mathsf R}}_g$ are 
possible corresponding to various groups which exist between ${\mathbb Z}_2$ and the group of rotations.

Instead of rotating the order parameter in a given magnetic field as in Eq.~(\ref{FT1}), we may fix the 
order parameter and rotate the magnetic field, which leads to:
\be
P_{\bf B}({\bf M}) = P_{\bf B'}({\bf M}) \ {\rm e}^{\beta ({\bf B}-{\bf B'})\cdot{\bf M}}, 
\label{FT2}
\ee
where ${\bf B'}={\boldsymbol{\mathsf R}}_g^{-1{\rm T}}\cdot\B$ for all $g\in G$.  
Furthermore, we notice that the fluctuation relations~(\ref{FT1}) and~(\ref{FT2}) hold exactly in finite systems.

If Eq.~(\ref{Jarzynski}) and Eq.~(\ref{Crooks}) may be considered as the equilibrium analogues of Jarzynski and Crooks identities, 
Eqs.~(\ref{FT1})-(\ref{FT2}) represent instead analogues of the Gallavotti-Cohen relation \cite{GC95}.

\subsection{Fluctuation-response relations}

Also in complete analogy with the non-equilibrium case, our fluctuation relation 
has implications for fluctuation-response relations. These are obtained  
 by expanding the left-hand side of Eq.~(\ref{Extension-gen-observable}) in powers of $\beta \B \cdot \M$ to first order:
 \be
\label{FDT1}
\langle A(\s)  \left[ 1 - \beta \B \cdot \M_N(\s) \right] \rangle_\B \simeq 
\frac{Z_N({\bf 0})}{Z_N({\B})} \langle A(\s) \rangle_{\bf 0} \, .
\ee
After taking the derivative with respect to $\B$ around $\B=0$ and expressing 
$\langle \M_N \rangle$ in terms of the partition function, one obtains
the fluctuation-response relation
\be
\label{FDT2}
\left. \frac{\partial\langle A(\s) \rangle}{\partial\B} \right|_{\B \rightarrow {\bf 0}} =\beta \left[ \langle A(\s) \,
\M_N(\s) \rangle_{\bf 0} 
- \langle A(\s) \rangle_{\bf 0} \, \langle \M_N(\s) \rangle_{\bf 0} \right] ,
\ee
which includes as a special case the well-known expression of the magnetic susceptibility in the direction $x$
$kT \chi=\langle M_{Nx}^2 \rangle_{\bf 0} - \langle M_{Nx} \rangle_{\bf 0}^2$, when $A(\s)=M_{Nx}(\s)$. For any finite-size system, the last term 
of Eq.~(\ref{FDT2}) vanishes since $\langle \M_N \rangle_{\bf 0}={\bf 0}$ in such a case. However, this is not necessarily 
the case if the thermodynamic limit is taken before the limit of $\B$ going to zero.  
When this happens, spontaneous symmetry breaking occurs as discussed in the next section.

By taking higher-order derivatives with respect to $\beta \B \cdot \M$ in Eq.~(\ref{Extension-gen-observable}), 
generalizations of the fluctuation-response relation can be obtained beyond the linear order.
Furthermore, this relation can be generalized to the case of an 
inhomogeneous magnetic field. By denoting $M_i$ the magnetization of 
the site~$i$ and $B_j$ the local magnetic field at the site~$j$ in some specific direction, one obtains 
another well-known expression for the magnetic susceptibility functional $\chi_{ij}$, 
which generalizes $\chi$ to spatially dependent magnetic field:
\be
\label{Fluct-response-gen}
\chi_{ij}=\left. \frac{\partial \langle M_i \rangle}{\partial B_j} \right|_{B \rightarrow 0} =\beta \left( \langle M_i M_j \rangle_0 
- \langle M_i \rangle_0 \langle M_j \rangle_0 \right).
\ee

\subsection{Inequalities implied by the isometric fluctuation relation}

Several inequalities can be deduced from the isometric fluctuation relation combined with the non-negativity of the Kullback-Leibler divergence:
\be
D[ P({\bf M}) || {\tilde P}({\bf M})]= \int d{\bf M} \, P({\bf M}) \, \ln \frac{P({\bf M})}{\tilde P({\bf M})} \geq 0. 
\label{Ineq-M}
\ee
Using $P({\bf M})=P_{\bf B}({\bf M})$ and ${\tilde P}({\bf M})=P_{\bf B}({\bf M'})$ with ${\bf M'}={\boldsymbol{\mathsf R}}_g^{-1}\cdot{\bf M}$, 
one obtains from Eq.~(\ref{FT1}) the inequality
\be
{\bf B}\cdot\langle{\bf M}\rangle_{\bf B} \geq {\bf B}\cdot{\boldsymbol{\mathsf R}}_g^{-1}\cdot\langle{\bf M}\rangle_{\bf B} \qquad\forall \ g\in G. 
\ee

Instead, using $P({\bf M})=P_{\bf B}({\bf M})$ and $\tilde P({\bf M})=P_{\bf 0}({\bf M})$, one obtains from Eq.~(\ref{Crooks}) the further inequality
\be
F_N({\bf B}) \geq F_N({\bf 0}) - {\bf B}\cdot\langle{\bf M}\rangle_{\bf B} \, ,
\ee
which also follows from the Jarzynski-like equality, Eq.~(\ref{Jarzynski}), using Jensen's inequality.
Conversely, a still further inequality can be obtained with the choice 
$P({\bf M})=P_{\bf 0}({\bf M})$ and $\tilde P({\bf M})=P_{\bf B}({\bf M})$ with the result
\be
F_N({\bf 0}) - {\bf B}\cdot \langle{\bf M}\rangle_{\bf 0}  \geq F_N({\bf B}) \, ,
\ee
which reduces to $\Delta F \geq 0$ given that $\langle{\bf M}\rangle_{\bf 0}=0$ for finite systems. 

\subsection{Relative entropy between symmetry related Gibbsian canonical distributions}

An interesting question is to compare two Gibbsian canonical distributions that are related by some symmetry~$g$ of the group \cite{G12JSM}: 
on the one hand, the canonical distribution~(\ref{canonical}) and, on the other hand, the symmetry related distribution 
$\mu_{\bf B}^g=\mu_{\bf B}(\s^g)= \mu_{\bf B}(\boldsymbol{\mathsf R}_g \cdot \s)$.

Now, the non-negative Kullback-Leibler divergence between these two Gibbsian distributions gives
\be
D(\mu_{\bf B}\Vert\mu_{\bf B}^g) \equiv {\rm tr} \, \mu_{\bf B} \ln\frac{\mu_{\bf B}}{\mu_{\bf B}^g} = 
\beta\, {\bf B}\cdot\left(\boldsymbol{\mathsf 1}-\boldsymbol{\mathsf R}_g^{-1}\right)\cdot\langle{\bf M}\rangle_{\bf B} \geq 0 \qquad\forall  \, g\in G
\label{Ineq1}
\ee
where the trace ${\rm tr}$ denotes the sum $\sum_{\s}$ over the microstates and $\langle\cdot\rangle_{\bf B}$ the statistical average over the Gibbsian state $\mu_{\bf B}$ in the presence of the external field $\bf B$.

If the group $G$ is finite and $\{\boldsymbol{\mathsf R}_g\}$ is a non-trivial irreducible representation of the group, the property
\be
\sum_{g\in G} \boldsymbol{\mathsf R}_g^{-1} = \sum_{g\in G} \boldsymbol{\mathsf R}_g = 0
\ee
holds, as proved in Ref.~\cite{H89}.  Accordingly, we get the general inequality:
\be
\sum_{g\in G} D(\mu_{\bf B}\Vert\mu_{\bf B}^g) = \sum_{g\in G} {\rm tr} \, \mu_{\bf B} \ln\frac{\mu_{\bf B}}{\mu_{\bf B}^g} = \vert G\vert \, \beta\, {\bf B}\cdot\langle{\bf M}\rangle_{\bf B} \geq 0 \, ,
\label{Ineq2}
\ee
where $\vert G\vert$ denotes the cardinal of the finite group.  If the group $G$ is continuous, $dv_G$ is its invariant measure normalized to unity, and still $\int_{G} \, dv_G\, \boldsymbol{\mathsf R}_g = 0$, the inequality reads:
\be
\int_{G} \, dv_G\, D(\mu_{\bf B}\Vert\mu_{\bf B}^g) =  \beta\, {\bf B}\cdot\langle{\bf M}\rangle_{\bf B} \geq 0 \, .
\label{Ineq3}
\ee

\subsection{Extension to quantum systems}

In quantum systems, the observables are given by Hermitian operators and only commuting observables can be measured simultaneously.  However, the three Cartesian components of the total magnetization do not commute between each other.  Indeed, the total magnetization is defined as
\be
\hat{\bf M}_N = \gamma \sum_{j=1}^{N} \hat{\bf S}_{j} \qquad\mbox{with} \quad \hat{\bf S}_{j}=(\hat S_{xj},\hat S_{yj},\hat S_{zj}) \, ,
\label{quant_magn}
\ee
in terms of the gyromagnetic ratio $\gamma$ and the non-commuting quantum spin operators $\{\hat{\bf S}_{j}\}_{j=1}^N$ such that
\be
[\hat S_{xj},\hat S_{yk}]=\mathrm{i}\, \hbar\, \hat S_{zj}\, \delta_{jk}\, ,
\label{spin_commutator}
\ee
where $\hbar$ is Planck's constant and $j,k=1,2,...,N$.  It is known that the operator giving the square of a spin vector is equal to $\hat{\bf S}_j^2=\hbar^2 s(s+1)\hat I$ where the spin quantum number $s$ may take integer or half-integer values: $s=0,\frac{1}{2},1,\frac{3}{2},2,...$ \cite{CDL91}.  
The correspondence between quantum and classical spins is established in the limit $s\to\infty$ by introducing the operators $\hat{\bf s}_j\equiv 
\hat{\bf S}_j/(\hbar\sqrt{s(s+1)})$, which have the following commutators: $[\hat s_{xj},\hat s_{yk}]=\mathrm{i}\, \hat s_{zj}\,\delta_{jk}/\sqrt{s(s+1)}$. 
In the limit $s\to\infty$, these commutators vanish so that the operators $\hat{\bf s}_j$ become the classical commuting variables $\s_j$ defining the unit 
vectors $\Vert\s_j\Vert=1$ that are the classical spins introduced in Subsection~\ref{IFR-begin}.  The classical magnetization~(\ref{Magnet}) is thus obtained
 by taking a gyromagnetic ratio such that $\gamma\hbar\sqrt{s(s+1)}=1$ in the limit $s\to\infty$.

As the consequence of Eqs.~(\ref{quant_magn})-(\ref{spin_commutator}), the commutator between two components of the magnetization does not vanish
\be
[\hat M_{Nx},\hat M_{Ny}]  = \mathrm{i} \, \hbar \, \gamma \, \hat M_{Nz} \, ,
\ee
and it is impossible to define a multivariate probability distribution for the vectorial magnetization, 
except in the limit $s \to \infty$ where the commutator vanishes and the classical limit should be recovered.

In order to circumvent  this essential difficulty, we may introduce a unit vector $\bf n$ and define the magnetization in its direction as
\be
\hat M \equiv {\bf n}\cdot\hat{\bf M}_N \, .
\ee
This operator is Hermitian and its orientation $\bf n$ can be changed to probe the rotational symmetry of the system properties. 

The statistical average $\langle\hat X\rangle_{\B}={\rm tr}(\hat\rho_N \hat X)$ is carried out over the quantum canonical density operator
\be
\hat\rho_N = \frac{1}{Z_N(\B)}\, {\rm e}^{- \beta \hat H_N(\B)}
\label{quantum canonical}
\ee
where $\hat H_N(\B)=\hat H_N({\bf 0})-\B\cdot\hat{\bf M}_N$ is the Hamiltonian operator, and $Z_N(\B)={\rm tr}\, 
{\rm e}^{-\beta \hat H_N(\B)}$ is the quantum partition function such that ${\rm tr}\,\hat\rho_N =1$.

In the canonical state of Eq.~(\ref{quantum canonical}), the univariate probability density of the eigenvalues of the operator $\hat M$ is defined as
\be
P_{\B}(M;{\bf n}) \equiv \left\langle \delta(M-{\bf n}\cdot\hat{\bf M}_N)\right\rangle_{\B} = {\rm tr} \left[ \hat \rho_N \, 
\delta(M-{\bf n} \cdot\hat{\bf M}_N) \right],
\label{quant_prob}
\ee
which is a function of the real variable $M\in{\mathbb R}$, depending parametrically on the orientation ${\bf n}\in{\mathbb S}^2$. 
This probability density is normalized as
\be
\int_{-\infty}^{+\infty} dM \, P_{\B}(M;{\bf n}) = 1
\ee
for every unit vector ${\bf n}$.

Let us suppose that the Hamiltonian operator $\hat H_N({\bf 0})$ is invariant under a symmetry group $G$.  In the Hilbert space of the quantum system, 
the elements $g$ of the group act by unitary operators $\hat U_g$.  The invariance of the Hamiltonian is expressed by
\be
[\hat U_g, \hat H_N({\bf 0})] = 0 \qquad \forall g\in G \, .
\label{commut-U-H0}
\ee
On the other hand, considering the group $G=SO(3)$, the magnetization would be transformed as
\be
\hat U_g^{-1} \, \hat{\bf M}_N \, \hat U_g = {\boldsymbol{\mathsf R}}_g\cdot\hat{\bf M}_N \qquad \forall g\in G 
\label{commut-U-MN}
\ee
with ${\boldsymbol{\mathsf R}}_g\in$~SO(3).

Since rotations are generated in quantum mechanics by the total angular momentum $\hat{\bf S}_{\rm tot}
\equiv\sum_{j=1}^{N} \hat{\bf S}_{j}$ and the magnetization is defined by Eq.~(\ref{quant_magn}), the unitary operator 
corresponding to a rotation of angle $\theta_g$ 
around the axis pointing in the direction of the unit vector ${\bf u}_g$ is given by
\be
\hat U_g =\exp\left(-\mathrm{i}\,\frac{\theta_g}{\hbar}\, \hat{\bf S}_{\rm tot} \cdot{\bf u}_g\right) .
\ee
Since $\hat{\bf M}_N=\gamma \hat{\bf S}_{\rm tot}$, the Hamiltonian operator $\hat H_N({\bf 0})$ commutes with the total magnetization, 
$[\hat{\bf M}_N, \hat H_N({\bf 0})] = 0$, so that there is the factorization ${\rm e}^{- \beta \hat H_N(\B)}={\rm e}^{- \beta \hat H_N({\bf 0})}{\rm e}^{\beta {\bf B}\cdot\hat{\bf M}_N}$.  However, we have that
\be
[{\bf B}\cdot\hat{\bf M}_N,{\bf n}\cdot\hat{\bf M}_N] = \mathrm{i} \, \hbar \, \gamma \, ({\bf B}\times{\bf n})\cdot\hat{\bf M}_N \, ,
\label{commut-B-n}
\ee
which vanishes only and only if the magnetic field $\bf B$ is parallel to the arbitrary direction $\bf n$.  Consequently, if we choose this direction along the magnetic field, ${\bf B}=B{\bf n}$, we obtain:
\be
P_{B{\bf n}}(M;{\bf n})  = \frac{Z_N({\bf 0})}{Z_N({\bf B})}\; {\rm e}^{\beta B M} \; P_{\bf 0}(M;{\bf n}) \, ,
\label{QGI}
\ee
which is the quantum version of Eq.~(\ref{GI}).

In the general case where the magnetic field $\bf B$ is not parallel to the direction $\bf n$, it is nevertheless possible to insert the 
identity relation $\hat U_g^{-1} \, \hat U_g =\hat I$ inside the definition of the probability density~(\ref{quant_prob}) and use 
Eqs.~(\ref{commut-U-H0})-(\ref{commut-U-MN}) to obtain the isometric fluctuation relation in the quantum framework as
\be
P_{\B}(M;{\bf n})=P_{{\boldsymbol{\mathsf R}}_g^{\rm T}\cdot\B}(M;{\boldsymbol{\mathsf R}}_g^{\rm T}\cdot{\bf n}) \qquad \forall g\in G \, .
\ee
Contrary to the classical results given by Eq.~(\ref{FT1}) or~(\ref{FT2}), exponential factors may not be moved outside the probability densities 
because the commutator~(\ref{commut-B-n}) does not vanish in quantum mechanics.

\section{Isometric fluctuation relations in infinite systems}
\label{IFRinf}

\subsection{Large-deviation functions of the order parameter}
\label{IFRinf-LD}

In the infinite-system limit $N\to\infty$, the fluctuation relations have their counterparts in terms of 
large-deviation functions \cite{E85,E95,D07,T09}. By defining the magnetization per spin $\m=\M/N$, one can 
introduce a large-deviation function $\Phi_{\bf B}(\m)$ such that
\be
P_{\bf B}(N{\bf m}) = A_N(\m)\, {\rm e}^{- N \Phi_{\bf B}(\m)} 
\label{LDP}
\ee
where $A_N(\m)$ is a prefactor which has a negligible contribution to $\Phi_{\bf B}(\m)$ 
in the limit $N \rightarrow \infty$.  As a result, Eq.~(\ref{FT1}) implies the following symmetry relation for the large-deviation function:
\be
\Phi_{\bf B}(\m) - \Phi_{\bf B}({\bf m'}) = \beta \B \cdot \left( {\bf m'} - \m \right)
\label{FT-LDP}
\ee
with $\Vert{\bf m'}\Vert=\Vert{\bf m}\Vert$.
It is important to appreciate that the function $\Phi_{\bf B}(\m)$ characterizes 
the equilibrium fluctuations of the order parameter, 
which are in general non Gaussian.

Using Eqs.~(\ref{GI}) and~(\ref{LDP}), this function can be expressed as
\be
\Phi_{\bf B}(\m)=\Phi_{\bf 0}(\m)-\beta\, \B \cdot \m - \beta  f(\B) + \beta f({\bf 0})\, ,
\ee
in terms of the Helmholtz free energy per spin, $f(\B)=-\beta^{-1} \ln Z_N(\B)/N$. One can then show
that $f({\bf 0}) - f(\B)$ is the Legendre-Fenchel transform \cite{T09} of $\beta^{-1} \Phi_{\bf 0}(\m)$ according to
\bea
&&  f(\B) = {\rm Inf}_{\m} \left[ \beta^{-1} \Phi_{\bf 0}(\m) -  \B \cdot \m + f({\bf 0}) \right] , \\
&& \Phi_{\bf 0}(\m) = \beta\,  {\rm Sup}_{\B} \left[ f(\B)+ \B \cdot \m - f({\bf 0}) \right] ,
\eea
according to Eqs.~(\ref{normalization}) and~(\ref{LDP}) with $\B={\bf 0}$ \cite{G12JSM}.

Assuming that derivatives with respect to $\m$ or $\B$ exist, 
the most probable value of the magnetization can be derived from the condition
\be
\label{most-prob-m}
\frac{\partial  \Phi_{\bf B}(\m)}{\partial \m} = \frac{\partial  \Phi_{\bf 0}(\m)}{\partial \m}  - \beta\, \B  =0\, ,
\ee
 which defines $\m=\m_*(\B)$. As a result of the large-deviation structure, the alternative condition 
\be
\label{most-prob-m2}
\frac{\partial  \Phi_{\bf B}(\m)}{\partial \B} =-\beta \, \m - \beta\, \frac{\partial  f(\B)}{\partial \B} =0
\ee
 defines an equivalent relation $\B=\B_*(\m)$. Unlike the Helmholtz free energy $f(\B)$ 
or the related function $\Phi_{\bf 0}(\m)$, it is important to emphasize that the large-deviation 
function $\Phi_{\bf B}(\m)$ depends on both thermodynamically 
conjugated variables $\m$ and $\B$.

Regarding this point, it is interesting to mention that historically 
a thermodynamic function similar to $\Phi_{\bf B}(\m)$ (but without its modern 
interpretation in terms of fluctuations) 
had been introduced a long time before large-deviation theory was even invented. 
For instance, in the chapter on dielectrics of the volume of electrodynamics of 
continuous media by Landau and Lifshitz~\cite{LL60}, 
a thermodynamic function of both the electric field ${\bf E}$ and of the displacement vector ${\bf D}$ 
(and similarly a related one for magnetic systems in terms of the vectors $\B$ and $\M$) 
had been introduced with essentially such properties.

Furthermore, since the most probable value of $\m$ namely $\m_*$ satisfies $\Phi_{\bf B}(\m_*)=0$, 
one can use Eqs.~(\ref{most-prob-m})-(\ref{most-prob-m2}) to rewrite $\Phi_{\bf B}(\m)$ in the two following equivalent forms
\ba
\label{equiv_form_1}
\Phi_{\bf B}(\m) &=& \Phi_{\bf 0}(\m)-\Phi_{\bf 0}(\m_*) - \frac{\partial  \Phi_{\bf 0}(\m_*)}{\partial \m}  \cdot \left( \m - \m_* \right), \\ 
                         &=&\beta f(\B_*) -\beta f(\B) +  \beta\, \frac{\partial  f(\B_*)}{\partial \B} \cdot \left( \B - \B_* \right), 
                         \label{equiv_form_2}
 \ea
 which describe respectively large fluctuations of $\m$ at fixed $\B$, or of $\B$ at fixed $\m$.
 These relations take the form of a truncated Taylor expansion with respect to the most probable point,
either given in terms of $\m$ or of $\B$. 
 Note that a similar form has been obtained for the equilibrium large-deviation function 
 of density fluctuations derived in 
 Ref.~\cite{D07}. There is however one important difference, namely that the result of this reference
requires short-range interactions in order to neglect some surface terms, while there is no
assumption of this kind to derive Eqs.~(\ref{equiv_form_1})-(\ref{equiv_form_2}).
Given Eq.~(\ref{most-prob-m}), the relation~(\ref{equiv_form_1}) can be rewritten as
 \be
 \Phi_{\bf B}(\m) = \Phi_{\bf 0}(\m)-\Phi_{\bf 0}(\m_*) + \beta \B \cdot \left( \m - \m_* \right),
 \ee
 which shows that the fluctuation relation~(\ref{FT-LDP}) immediately holds when 
$\Phi_{\bf 0}(\m)=\Phi_{\bf 0}(\boldsymbol{\mathsf R}_g^{-1}\cdot\m)$.

\subsection{Implications for spontaneous broken symmetry}

In order to discuss spontaneously broken symmetry, let us introduce the cumulant generating function for the magnetization:
\be
\Gamma_{\bf B}(\pmb{\lambda}) \equiv \lim_{N\to\infty}-\frac{1}{N}\ln \left\langle{\rm e}^{-\pmb{\lambda}\cdot{\bf M}_N(\pmb{\sigma})}\right\rangle_{\bf B} ,
\label{GF}
\ee
which is the Legendre-Fenchel transform of the function $-\Phi_{\bf B}(\m)$ defined by Eq.~(\ref{LDP}):
\be
\Gamma_{\bf B}(\pmb{\lambda})  = {\rm Inf}_{\m} \left[ \Phi_{\bf B}(\m) +  \pmb{\lambda} \cdot \m \right] \, .
\ee

As a consequence of the isometric fluctuation relation~(\ref{FT1}), 
the generating function~(\ref{GF}) obeys the symmetry relation
$\Gamma_{\bf B}(\pmb{\lambda}) = \Gamma_{\bf B}\left[\beta{\bf B}+{\boldsymbol{\mathsf R}}_g^{\rm T}\cdot(\pmb{\lambda}-\beta{\bf B})\right]$ 
for all $g\in G$.  In the particular case of the inversion ${\boldsymbol{\mathsf R}}_g=-{\boldsymbol{\mathsf 1}}$, 
this symmetry relation reduces to
\be
\Gamma_{\bf B}(\pmb{\lambda}) = \Gamma_{\bf B}(2\beta{\bf B}-\pmb{\lambda})\, ,
\label{FT-GF}
\ee
whereupon the average magnetization per spin, which is equal to the first cumulant, is given by
\be
\langle{\bf m}\rangle_{\bf B} = \frac{\partial \Gamma_{\bf B}}{\partial\pmb{\lambda}}({\bf 0}) = -\frac{\partial \Gamma_{\bf B}}{\partial\pmb{\lambda}}(2\beta{\bf B}) \, .
\label{FT-GF-M}
\ee
This result is of fundamental importance for SSB.
Indeed, as long as the cumulant generating function~(\ref{GF}) remains analytic in the variables $\pmb{\lambda}$
(which is necessarily the case in a finite system), the average magnetization has to vanish in the absence of external field
because $\langle{\bf m}\rangle_{\bf 0} = \partial_{\pmb{\lambda}}\Gamma_{\bf 0}({\bf 0}) = 
-\partial_{\pmb{\lambda}}\Gamma_{\bf 0}({\bf 0})=-\langle{\bf m}\rangle_{\bf 0}=0$, as implied by Eq.~(\ref{FT-GF-M}).

However, the generating function may no longer be analytic in the thermodynamic limit $N\to\infty$, allowing
a spontaneous magnetization $\langle{\bf m}\rangle_{\bf 0}\neq 0$ in the absence of external field.
At the critical temperature $T_c$ and near $\pmb{\lambda}=\beta{\bf B}$, the generating function has the universal scaling behavior
\be
\big\vert \Gamma_{\bf B}(\pmb{\lambda})-\Gamma_{\bf B}(\beta{\bf B})\big\vert_{T_c} \sim \Vert\pmb{\lambda}-\beta{\bf B}\Vert^{1+1/\delta}
\label{USB}
\ee
in accordance with the scaling $\Vert\langle{\bf m}\rangle_{\bf B}\Vert_{T_c}\sim\Vert{\bf B}\Vert^{1/\delta}$ for the critical magnetization.
The critical exponent takes the value $\delta=3$ in the mean-field models and $\delta=15$ in the two-dimensional Ising model \cite{W65,F67,KGHHLPRSAK67,H87}.  The non-analyticity of the cumulant generating function allows for the possibility of a non-vanishing spontaneous magnetization in the thermodynamic limit and thus of SSB.
This non-analytic behavior is in fact inherited from the non-analyticity of the free energy in the same conditions and is related to a well-known theorem proven by Lee and Yang \cite{LY52}.

\subsection{Fluctuation relation for a local vectorial order parameter}

The isometric fluctuation relation~(\ref{FT1}) also holds for a spatially
varying magnetic field $\B(\sr)$ and magnetization density ${\bf m}(\sr)=\sum_{i=1}^N \s_i\, \delta(\sr - \sr_i)$, 
where $\sr_i$ is the location of spin~$\s_i$.   In order to show this result, we need
to coarse grain the magnetization density.  By adapting the derivation of Eq.~(\ref{FT1}), one then finds
\be
P_{\bf B}[{\bf m}(\sr)]=  P_{\bf B}[{\bf m'}(\sr)] \, {\rm e}^{\beta \int d\sr \, \B(\sr) 
\cdot [{\bf m}(\sr)-{\bf m'}(\sr) ]} ,
\label{FT-Mofr}
\ee  
where $P_{\bf B}[{\bf m}(\sr)]$ is the probability functional of the magnetization density 
${\bf m}(\sr)$ and ${\bf m'}(\sr)={\boldsymbol{\mathsf R}}_g^{-1} \cdot{\bf m}(\sr)$ (see the Supplementary Material of Ref.~\cite{LG14} for detail).
Using this relation, it is possible to derive in an alternate way the isometric fluctuation relation of Eq.~(\ref{FT1}) under the assumption that 
the magnetic field is uniform, by defining
the probability density of the order parameter $\M$ as
\be
P_{\bf B}(\M)= \int \mathcal{D}{\bf m}(\sr) \, P_{\bf B}[{\bf m}(\sr)] \, \delta \left[ \M - \int d \sr \, {\bf m}(\sr) \right].
\ee
This represents an alternate route using a local formulation to obtain the isometric fluctuation relation of Eq.~(\ref{FT1}), 
which is similar in spirit to the approach followed in Refs.~\cite{HPPG11,VHT04} for the non-equilibrium case. 
Note however, that the isometric fluctuation relation of Eq.~(\ref{FT1}) are exact and do not require additional 
assumptions for their derivation, unlike their non-equilibrium counterparts which have been derived only within the framework
of the Macroscopic Fluctuation Theory.

The functional $P_{\bf B}[{\bf m}(\sr)]$ also satisfies a large-deviation principle of the form
\be
P_{\bf B}[{\bf m}(\sr)]= {\rm e}^{- \beta \int d \sr \, \omega_{\bf B}[{\bf m}(\sr), \nabla {\bf m}(\sr),...]} ,
\label{Prob-Omega}
\ee
where $\Omega_{\bf B}[{\bf m}(\sr)]=\int d \sr \, \omega_{\bf B}[{\bf m}(\sr), \nabla {\bf m}(\sr),...]$ represents the extensive Ginzburg-Landau functional associated with a coarse-grained magnetization density field ${\bf m}(\sr)$.  
This functional carries with it a cutoff length associated with the coarse-grained order parameter.
Within the Ginzburg-Landau approach, only the extremals of the functional $\Omega_{\bf B}[{\bf m}(\sr)]$ with respect to ${\bf m}(\sr)$ are kept, an approximation which is analyzed by the so-called Ginzburg-Landau criterium \cite{CL95}. However, the functionals $P_{\bf B}[{\bf m}(\sr)]$ and $\Omega_{\bf B}[{\bf m}(\sr)]$ contain all the information about the fluctuations of the order parameter.

Now, substituting Eq.~(\ref{Prob-Omega}) into Eq.~(\ref{FT-Mofr}), we obtain
\be
\Omega_{\B} [{\bf m}(\sr)] = \Omega_{\B} [{\bf m'}(\sr)] - \beta \int d\sr \, \B(\sr) 
\cdot [{\bf m}(\sr)-{\bf m'}(\sr) ] \, .
\label{Omega-Mofr}
\ee 
We note that such a relation is less general than its counterpart in Eq.~(\ref{FT-Mofr}) due to the 
requirement of a large-deviation principle for the probability distribution to establish Eq.~(\ref{Omega-Mofr}). 
Furthermore, the large-deviation function for the magnetization density field, $\Omega_{\B} [{\bf m}(\sr)]$, can also be expressed as
\be
\Omega_{\B} [{\bf m}(\sr)] = \Omega_{\bf 0} [{\bf m}(\sr)] - \Omega_{\bf 0} [{\bf m}_*(\sr)] - \beta \int d\sr \, \B(\sr) 
\cdot [{\bf m}(\sr)-{\bf m}_*(\sr) ]
\ee
relative to the most probable and uniform magnetization density ${\bf m}_*(\sr)$, which is solution of
\be
\frac{\delta \Omega_{\B}}{\delta{\bf m}(\sr)}[{\bf m}_*(\sr)] = \frac{\delta \Omega_{\bf 0}}{\delta{\bf m}(\sr)}[{\bf m}_*(\sr)] -\B(\sr) = 0
\ee
in analogy with the results of Subsection~\ref{IFRinf-LD}.

\section{Applications to isotropic magnetic systems}
\label{Magnetic}

Several illustrative examples of isotropic magnetic systems are studied in this section. We first present solvable models such as the 3D Curie-Weiss model and 
the 1D Heisenberg chain. We then consider the XY model mainly numerically.

\subsection{The three-dimensional Curie-Weiss model}

Let us start by considering $N$ classical spins $\{\s_i\}_{i=1}^N$ defined by unit vectors $\Vert\s_i\Vert=1$ with the mean-field 
Curie-Weiss interaction. The Hamiltonian of this system is given by
\be
H_N(\pmb{\sigma};{\bf B})=-\frac{J}{2N} \, {\bf M}_N(\pmb{\sigma})^2 -{\bf B}\cdot{\bf M}_N(\pmb{\sigma})\, .
\ee
The interaction is long-ranged so that the coupling is global between all the spins.
The probability distribution of the order parameter can be expressed as
\be
P_{\bf B}({\bf M})=\frac{1}{Z_N({\bf B})} \ {\rm e}^{\frac{\beta J}{2N}\, {\bf M}^2 +\beta{\bf B}\cdot{\bf M}}  C_N({\bf M})\, , 
\label{P(M)}
\ee
in terms of the function
\be
C_N({\bf M}) \equiv \int \frac{d^N\pmb{\sigma}}{(4\pi)^N} \ \delta\left[{\bf M}-{\bf M}_N(\pmb{\sigma})\right] 
\label{fn_C}
\ee
representing the number of microstates with a given magnetization $\M$. This number, which is rotationally invariant, is related by $C_N={\rm e}^{S_N/k}$ to the entropy function $S_N(\M)$ and Boltzmann's constant $k$. Using large-deviation theory \cite{LG14,E85,E95,T09}, one explicitly obtains this entropy in the form of $S_N(N\m)=-k N I(m)$ with $m=\Vert\m\Vert$ and
\be
I(m)=m \mathcal{L}^{-1} (m)- \ln \frac{\sinh\left[\mathcal{L}^{-1}(m)\right]}{\mathcal{L}^{-1}(m)},
\label{I(m)}
\ee
where $\mathcal{L}^{-1}$ is the inverse of the Langevin function $\mathcal{L}(x)=\coth(x)-1/x$, 
a result which also follows from a standard mean-field approach \cite{LL01}. 
Combining Eqs.~(\ref{P(M)})-(\ref{I(m)}), the large-deviation function defined in Eq.~(\ref{LDP}) is obtained as \cite{Footnote}
\be
\Phi_\B(\m) = I(m) - \beta J m^2/2 - \beta \B \cdot \m - \beta f(\B) \, .
\ee 
The calculation of the prefactor $A_N(\m)$ of Eq.~(\ref{LDP}) is provided in the Supplementary Material of Ref. \cite{LG14},
together with an alternate derivation of the entropy function $S_N(\M)$, based on the property that 
this function is also the Shannon entropy of the angular distribution of the spins, which can be calculated exactly for this model.

\begin{figure}[h!]
\includegraphics[scale=0.45]{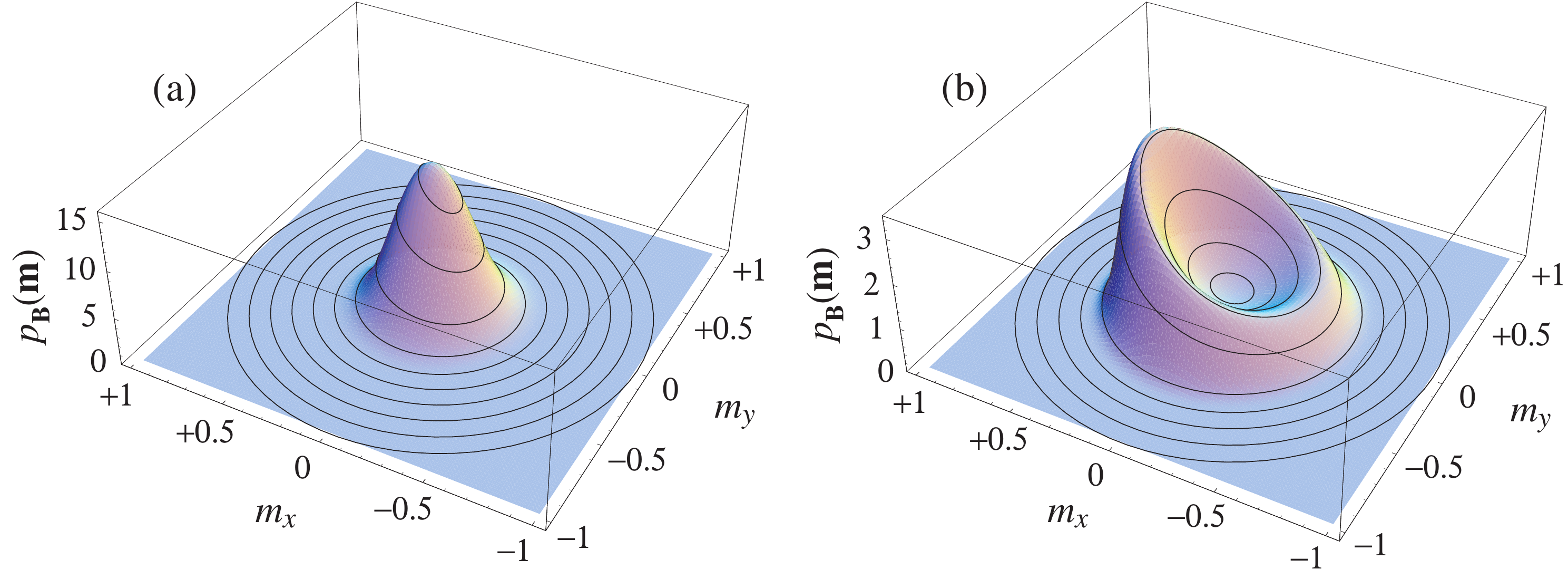}
\caption{Probability density $p_\B(\m)=N^3 P_\B(N\m)$ of the magnetization per spin $\m=\M/N=(m_x,m_y,m_z=0)$ for the three-dimensional Curie-Weiss model in the magnetic field $\B=(B,0,0)$ with $B=0.005$, $J=1$, and $N=100$ at the rescaled inverse temperatures (a) $\beta J=2.7$ in the paramagnetic phase and (b) $\beta J=3.3$ in the ferromagnetic phase.  The lines depict the contours of $\Vert\m\Vert=0.1,0.2,...,1.0$ where the isometric fluctuation relation (\ref{FT1}) holds.}
\label{fig1}
\end{figure}

\begin{figure}[h]
\includegraphics[scale=0.5]{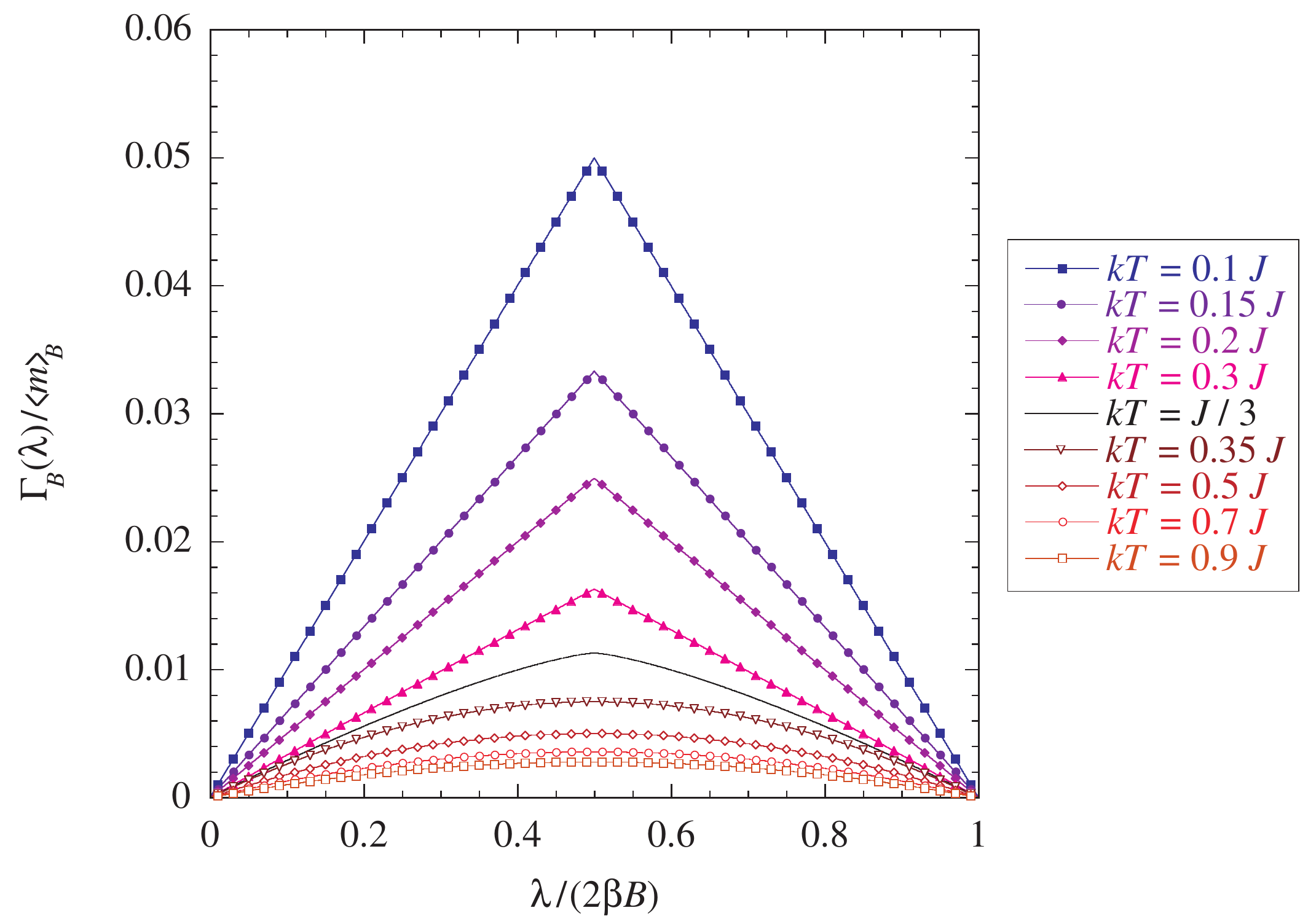}
\caption{Cumulant generating function (\ref{GF}) of the magnetization in the three-dimensional Curie-Weiss model for the magnetic field $\B=(B,0,0)$ with $B=0.005$, $J=1$, and different temperatures across criticality.  The generating function is rescaled by the average magnetization $\langle m\rangle_B$ in the direction of the external field and plotted versus the rescaled parameter $\lambda/(2\beta B)$ in the same direction $\pmb{\lambda}=(\lambda,0,0)$.  The generating function is computed by taking the Legendre-Fenchel transform of the large-deviation function $\Phi_{\bf B}({\bf m})$ introduced in Eq.~(\ref{LDP}).  The isometric fluctuation relation (\ref{FT1}) implies the symmetry $\lambda\to2\beta B-\lambda$ of the generating function according to Eq.~(\ref{FT-GF}).}
\label{fig2}
\end{figure}

For this model, it is straightforward to check that this large-deviation function satisfies the symmetry relation of Eq.~(\ref{FT-LDP}).
In Fig.~\ref{fig1}, the probability density $p_{\bf B}({\bf m})=N^3 P_{\bf B}(N{\bf m})$ is depicted as a function of the  components $(m_x,m_y)$ of the magnetization per spin. Above the critical temperature $kT_c=J/3$, the distribution presents a maximum close to the 
origin in Fig.~\ref{fig1}a, while the symmetry breaking manifests itself by a crater-like distribution below the critical temperature in Fig.~\ref{fig1}b. 
The presence of the external magnetic field ${\bf B}=(0.005,0,0)$ tends to shift the distribution in the same direction in the magnetization space, as seen in 
Fig.~\ref{fig1}.  Now, if ${\bf B}=(B,0,0)$, the fluctuation relation~(\ref{FT1}) can be written for this density as
\be
p_{\bf B}(m\cos\theta,m\sin\theta,0)=p_{\bf B}(m,0,0)\, {\rm e}^{N\beta B m (\cos\theta-1)}
\ee
along the lines at constant values of $m=\sqrt{m_x^2+m_y^2}$.  As observed in Fig.~\ref{fig1}, these lines coincide with the surface of the distribution in agreement with the isometric fluctuation relation.

Figure~\ref{fig2} shows the cumulant generating function~(\ref{GF}) below and above the critical temperature.  This function is analytic in the paramagnetic phase above the critical temperature, 
but this is no longer the case below the critical temperature in the ferromagnetic phase where the function becomes non-analytic at the symmetry point $\pmb{\lambda}=\beta{\bf B}$.  As aforementioned, 
this non-analyticity is at the origin of the spontaneous magnetization in the ferromagnetic phase.  At the critical temperature, the generating function has the universal scaling behavior~(\ref{USB}) 
with $\delta=3$, as it should for this mean-field model \cite{LG14}.  The remarkable result is that the symmetry $\lambda\to2\beta B-\lambda$ of the isometric fluctuation relation remains satisfied across the phase transition.

Many systems with long-range interactions can be treated like this Curie-Weiss model. For more examples of the use of large-deviation theory for such models, 
we refer to the review \cite{CDR09}.

\subsection{The one-dimensional Heisenberg chain}

Besides the mean-field Curie-Weiss model, the isometric fluctuation relation applies as well to magnetic systems with short-range interaction. In particular, the relation can be established using the transfer-matrix method which is
 applicable to the case of a one-dimensional chain of classical spins in a magnetic field \cite{BHL75}, as shown below. In the absence of external field, the Hamiltonian of this one-dimensional Heisenberg model is
\be
H_0 = - J \sum_{i=1}^N \s_i\cdot\s_{i+1}
\label{H0-1d-chain}
\ee
where $\s_i\in{\mathbb S}^2$ are unit vectors on the sphere. The Hamiltonian~(\ref{H0-1d-chain}) is symmetric under the rotations of SO(3) and, more generally, 
the orthogonal transformations of O(3). The coupling to the external magnetic field is described by the Hamiltonian~(\ref{Hamilt}) with the 
magnetization~(\ref{Magnet}).

The cumulant generating function is thus defined by Eq.~(\ref{GF}) where
\be
\langle{\rm e}^{-\pmb{\lambda}\cdot{\bf M}_N}\rangle_{\bf B}=\frac{{\rm tr}\left(\hat V _{{\bf B},\pmb{\lambda}}\right)^N}{{\rm tr}\left(\hat V _{{\bf B},{\bf 0}}\right)^N}
\ee 
can be expressed in terms of the transfer operator \cite{BHL75}, which is such that for any function $\psi(\s)$
\be
\hat V _{{\bf B},\pmb{\lambda}}\psi(\s) \equiv \int d\s' \, V _{{\bf B},\pmb{\lambda}}(\s\vert\s') \, \psi(\s')
\ee
where the integral kernel is defined by
\be
V _{{\bf B},\pmb{\lambda}}(\s\vert\s')=\exp\left[ \beta J \, \s\cdot\s' + (\beta{\bf B}-\pmb{\lambda})\cdot\frac{\s+\s'}{2}\right] .
\label{V-dfn}
\ee
This transfer operator has the symmetry
\be
\hat R_g^{-1}\,\hat V _{{\bf B},\pmb{\lambda}}\, \hat R_g=
\hat V _{{\bf B},\boldsymbol{\mathsf R}_g\cdot(\pmb{\lambda}-\beta{\bf B})+\beta{\bf B}}
\label{V-sym}
\ee
where $\hat R_g \psi(\s)=\psi(\boldsymbol{\mathsf R}_g\cdot\s)$
and $\boldsymbol{\mathsf R}_g$ is the $3\times 3$ matrix representing the group element $g\in$~O(3).

As a consequence, the cumulant generating function has the symmetry:
\be
\Gamma_{\bf B}(\pmb{\lambda})=
\Gamma_{\bf B}\left[\boldsymbol{\mathsf R}_g\cdot(\pmb{\lambda}-\beta{\bf B})+\beta{\bf B}\right]
\label{G-sym}
\ee
of the isometric fluctuation relation.

\begin{figure}[h]
\includegraphics[scale=0.5]{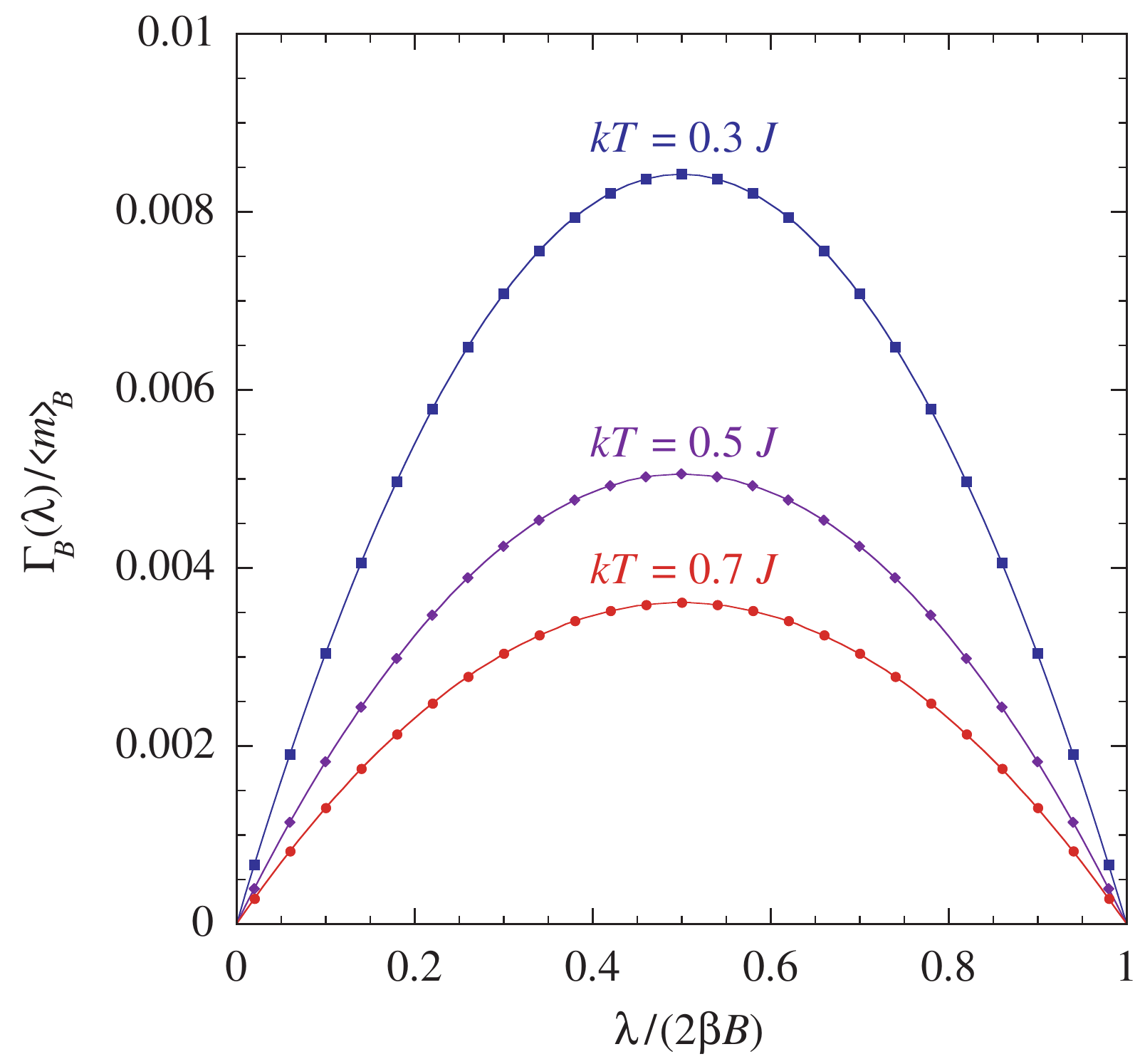}
\caption{Cumulant generating function (\ref{GF}) of the magnetization in the one-dimensional Heisenberg chain for the magnetic field $\B=(0,0,B)$ with $B=0.005$, $J=1$, and different temperatures.  The generating function is rescaled by the average magnetization $\langle m\rangle_B$ in the direction of the external field and plotted versus the rescaled parameter $\lambda/(2\beta B)$ in the same direction $\pmb{\lambda}=(0,0,\lambda)$.  The symmetry $\lambda\to2\beta B-\lambda$ of the generating function is the test of Eq.~(\ref{FT-GF}) implied by the isometric fluctuation relation (\ref{FT1}).}
\label{fig3}
\end{figure}

In order to test this relation, the cumulant generating function has been obtained numerically by iterating the transfer operator~(\ref{V-dfn}) starting from an initial function $\psi_0(\s)=1$. 
The unit sphere $\s=(\sin\theta\cos\phi,\sin\theta\sin\phi,\cos\theta)$ is discretized by using the variables $-1<\xi=\cos\theta\leq +1$ and $0\leq\phi< 2\pi$, in terms of which the element of integration
 is uniform: $d\s=d\cos\theta\,d\phi=d\xi \,d\phi$.  The $K^2$ grid points are spaced by $\Delta\xi=2/K$ and $\Delta\phi=2\pi/K$.  The transfer operator is iterated up to $N=100$ at different values of
 $\lambda$ to obtain approximations for the cumulative generating function converging as $1/N$.  The value for an infinite chain can be extrapolated by using this scaling.  Moreover, the convergence 
with $K$ scales as $1/K^2$, which can also be used to get the final value by extrapolation from calculations with $K=10,15,20$, and $25$.  The results are shown in Fig.~\ref{fig3} for different values of the 
temperature.  We note that there is no phase transition in one-dimensional systems so that the generating function remains analytical at positive values of the temperature.  The symmetry $\lambda\to 2\beta B-\lambda$
 of the fluctuation relation is confirmed.

 \subsection{The two-dimensional $XY$ model}

The isometric fluctuation relation also applies to the more complex $XY$~model \cite{CL95,KT73,BHP98,PHSB01,TC79,M84}. The Hamiltonian of the $XY$~model in an external magnetic field ${\bf B}=(B_x,B_y)$ is given by
\be
H_N(\pmb{\theta};{\bf B})= - J\, \sum_{\langle i,j\rangle} \cos(\theta_i-\theta_j) - {\bf B}\cdot{\bf M} \, ,
\label{H-XY}
\ee
on a square lattice with $N=L\times L$ sites ($i,j=1,2,...,N$) with the magnetization 
\be
{\bf M} = \sum_i \left(\cos\theta_i,\sin\theta_i\right),
\label{Magnet-XY}
\ee  
and $0 \leq \theta_i < 2 \pi$.
In the absence of external field, the Hamiltonian is symmetric under the orthogonal group O(2).  In view of Eq.~(\ref{GI}), the probability 
distribution of the magnetization in the field can be obtained from the same distribution in the absence of the field $P_{\bf 0}({\bf M})$. 
This quantity is itself related to the probability distribution of the modulus of the magnetization $Q(M)$ by  $P_{\bf 0}({\bf M})=Q(M)/(2 \pi M)$. 
The distribution $Q(M)$ has been observed in several contexts \cite{BHP98} and it has been shown to be numerically very close to a Gumbel 
distribution below the Kosterlitz-Thouless transition temperature~$T_{\rm KT}$ \cite{PHSB01}.

\begin{figure}[h]
\centerline{\scalebox{0.4}{\includegraphics{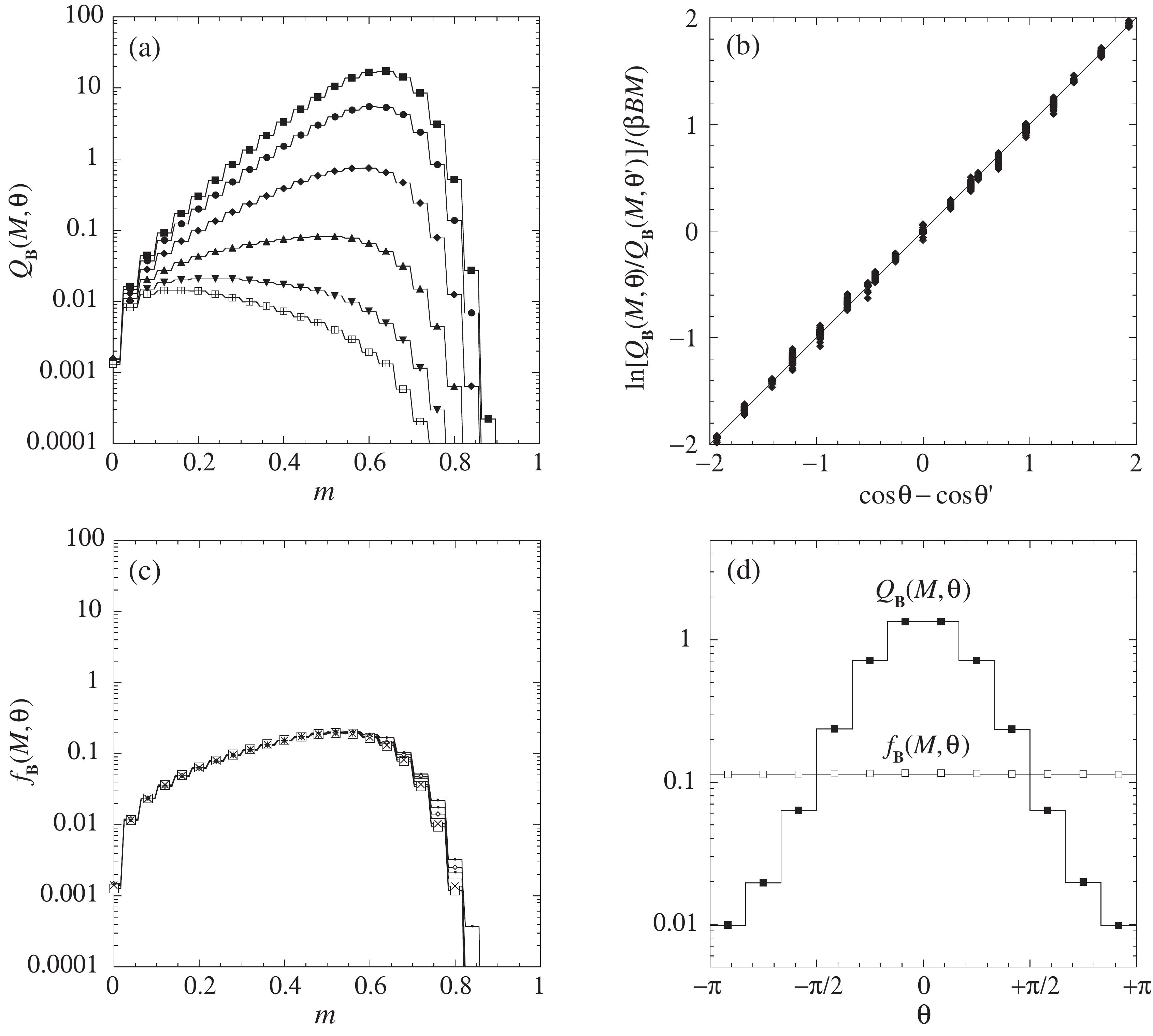}}}
\caption{$XY$ model on a square lattice with $L=10$ in an external magnetic field $(B_x=0.1,B_y=0)$ at the rescaled temperature $kT/J=1.25$:  
(a) The probability distributions (\ref{Q-XY}) of the magnetization per spin $m=M/L^2$.  (b) The rescaled logarithm (\ref{log-FR}) of the ratio
 of the probability distributions for the magnetizations $\bf M$ and $\bf M'$ such that $\Vert{\bf M}\Vert=\Vert{\bf M'}\Vert$ versus the 
difference of orientations between the magnetizations $\bf M$ and $\bf M'$ with respect to the external magnetic field $\bf B$ for values of 
the probability distribution function higher than $10^{-2}$.  (c) The distributions rescaled with the Boltzmann weights (\ref{f-XY}).  Their 
coincidence tests the fluctuation relation.  (d) The angular distributions without (filled squares) and with (open squares) the Boltzmann 
weights at the magnetization $m=0.32$.}
\label{fig4}
\end{figure}

\begin{figure}[h]
\centerline{\scalebox{0.4}{\includegraphics{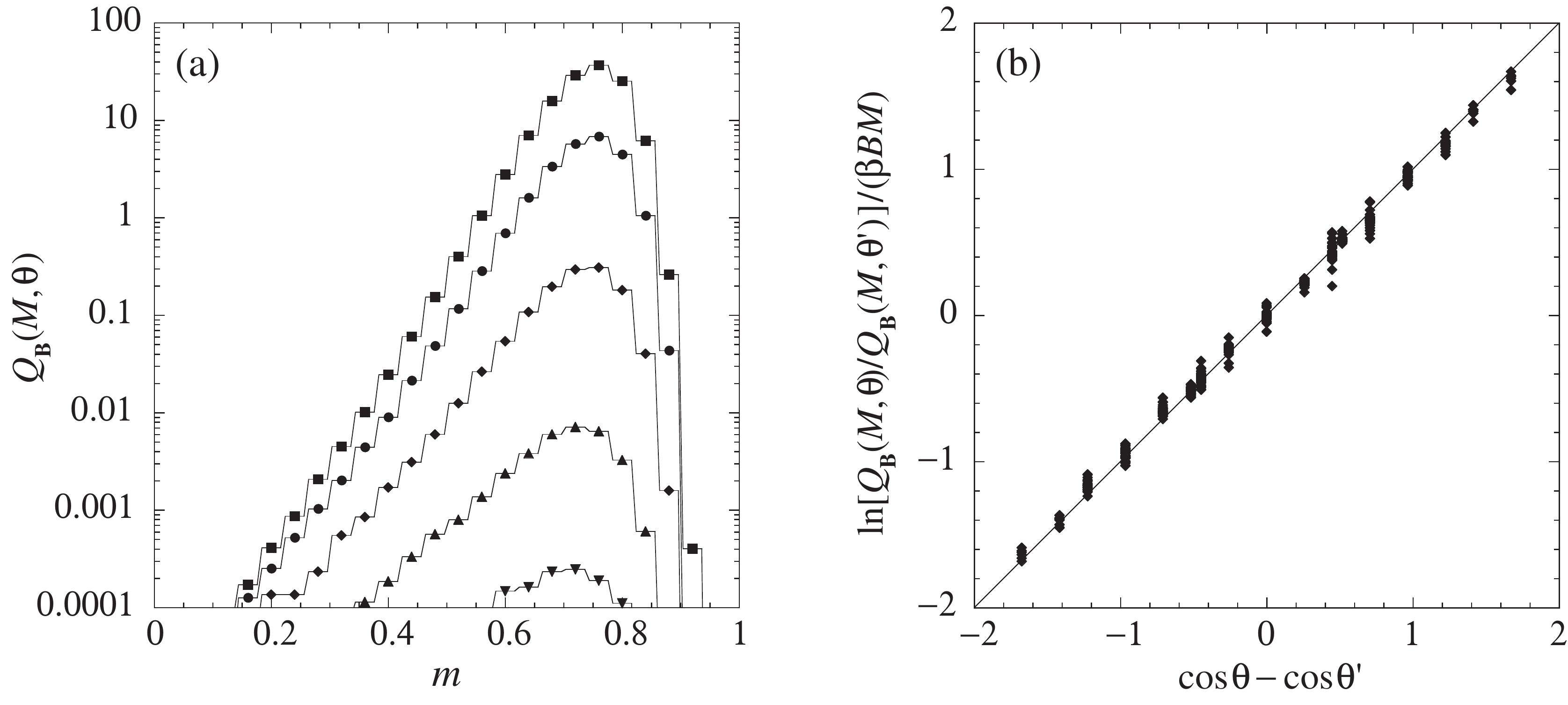}}}
\caption{$XY$ model on a square lattice with $L=10$ in an external magnetic field $(B_x=0.1,B_y=0)$ at the rescaled temperature $kT/J=1.00$:
  (a) The probability distributions (\ref{Q-XY}) of the magnetization per spin $m=M/L^2$.  (b) The rescaled logarithm (\ref{log-FR}) of the ratio 
of the probability distributions for the magnetizations $\bf M$ and $\bf M'$ such that $\Vert{\bf M}\Vert=\Vert{\bf M'}\Vert$ versus the 
difference of orientations between the magnetizations $\bf M$ and $\bf M'$ with respect to the external magnetic field $\bf B$ for values of 
the probability distribution function higher than $10^{-4}$.}
\label{fig5}
\end{figure}

\begin{figure}[h]
\centerline{\scalebox{0.4}{\includegraphics{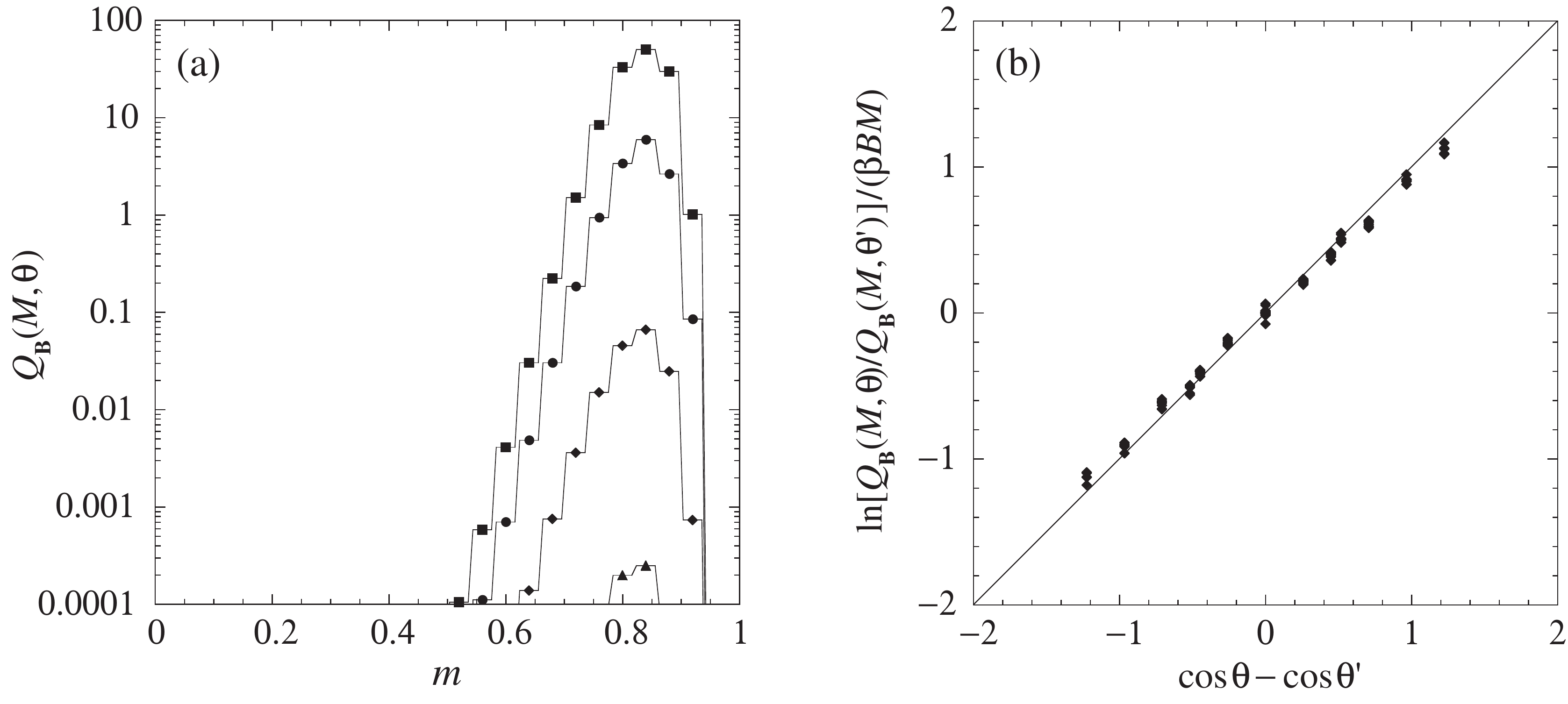}}}
\caption{$XY$ model on a square lattice with $L=10$ in an external magnetic field $(B_x=0.1,B_y=0)$ at the rescaled temperature $kT/J=0.75$:  (a) The probability distributions (\ref{Q-XY}) of the magnetization per spin $m=M/L^2$.  (b) The rescaled logarithm (\ref{log-FR}) of the ratio of the probability distributions for the magnetizations $\bf M$ and $\bf M'$ such that $\Vert{\bf M}\Vert=\Vert{\bf M'}\Vert$ versus the difference of orientations between the magnetizations $\bf M$ and $\bf M'$ with respect to the external magnetic field $\bf B$ for values of the probability distribution function higher than $10^{-5}$.}
\label{fig6}
\end{figure}

In order to test the isometric fluctuation relation, the canonical probability distribution
\be
Q_{\bf B}(M,\theta)\equiv 2 \pi M P_{\bf B}({\bf M})
\label{Q-XY}
\ee
has been obtained by Monte Carlo simulations.  This probability distribution is normalized as
\be
\int_0^{L^2} dM \int_0^{2\pi} \frac{d\theta}{2\pi} \, Q_{\bf B}(M,\theta) =1 \, .
\label{norm-Q-XY}
\ee
According to the Monte Carlo algorithm, a rotator $\theta_i$ is taken with equal probability on the $N=L^2$ sites of the lattice and is given a new orientation $\theta_i'$ if $\exp\left[-\beta\Delta E(\theta_i')\right]>u$ where $\Delta E(\theta_i')=H_{\bf B}(\pmb{\theta}')-H_{\bf B}(\pmb{\theta})$ is the energy difference due to the change of orientation and $u\in[0,1]$ is a uniformly distributed real random number. 
After a transitory run of $10^4$ spin flips, the statistics is carried out over $10^8$ values of the magnetization. 
Every value is sampled after $1000$ spin flips. 

Figure~\ref{fig4}a shows the distribution $Q_{\bf B}(M,\theta)$ versus versus the magnetization per spin $m=M/L^2$ for different values of the angle $\theta$. 
The histograms are established with $\Delta m=1/25$ and $\Delta\theta=\pi/6$.  The distribution is the largest in the direction of the external magnetic field ${\bf B}=(0.1,0)$.

Figure~\ref{fig4}b presents the test of the isometric fluctuation relation using an equivalent form put forward by Hurtado {\it et al.} \cite{HPPG11,HPPG14} 
\be
\frac{1}{\beta B M}\, \ln\frac{Q_{\bf B}(M,\theta)}{Q_{\bf B}(M,\theta')} = \cos\theta-\cos\theta' \, .
\label{log-FR}
\ee
We observe that the left- and right-hand sides of this relation are indeed equal within numerical errors, which confirms the isometric fluctuation relation.

Figure~\ref{fig4}c shows the probability distribution of the magnetization compensated with Boltzmann's weights:
\be
f_{\bf B}(M,\theta) \equiv Q_{\bf B}(M,\theta)\, {\rm e}^{-\beta B M \cos\theta} \, ,
\label{f-XY}
\ee
where $\theta$ is taken as the angle between the magnetization of magnitude $M$ and the external magnetic field of magnitude $B$.  We observe that these weighted distributions no longer depend on the angle $\theta$ and they collapse together in agreement with the isometric fluctuation relation.  Finally, the angular distributions without and with the Boltzmann weights are plotted in Fig.~\ref{fig4}d for the magnitude $m=0.32$ of the magnetization, confirming the absence of angular dependence for the weighted distribution, as expected from the isometric fluctuation relation.  

Figures~\ref{fig5} and~\ref{fig6} show the probability distribution of the magnetization and its symmetry at two lower temperatures than Fig.~\ref{fig4}, showing that the isometric fluctuation relation holds below the Kosterlitz-Thouless transition temperature $T_{\rm KT}\simeq 0.89 J/k$ \cite{TC79,M84}, as well as above.

\section{Applications to anisotropic systems}
\label{Anisotropic}

\subsection{General derivation}

In some situations, the physical system of interest is invariant under a discrete group instead of a continuous one, 
for instance when the system is anisotropic. 
In order to address such a case, let us consider a Hamiltonian $H_N(\s;{\bf 0})$ which is invariant
under the action of a group that may be discrete or continuous.

The fluctuation relation of Eq.~(\ref{FT1}) for the probability distribution 
that the order parameter would take the value ${\bf M}$ can be recast in the following way
\be
\frac{P_{\bf B}({\bf M})}{P_{\bf B}(\boldsymbol{\mathsf R}_g^{-1} \cdot{\bf M})} = 
\exp\left[ \beta\, {\bf B}\cdot\left(\boldsymbol{\mathsf 1}-\boldsymbol{\mathsf R}_g ^{-1} \right)\cdot{\bf M}\right] \qquad\forall  \, g\in G
\label{FR}
\ee
where ${\bf M}'$ has been replaced by $\boldsymbol{\mathsf R}_g^{-1} \cdot{\bf M}$.

For the group $G={\mathbb Z}_2$, we recover for $g=-{\rm Id}$ the previously obtained fluctuation relation, namely Eq.~(\ref{Goldenfeld}).  For the continuous group of rotations $G=$~SO(2) or $G=$~SO(3), we recover the isometric fluctuation relation already discussed in previous sections. The fluctuation relation (\ref{FR}) also holds for discrete groups in between ${\mathbb Z}_2$ and continuous groups, as illustrated with the following models.

\subsection{$q$-state Potts model}

The Hamiltonian of this other model is given by
\be
H_{0}= - \sum_{i<j} J_{ij} \, \delta_{\sigma_i,\sigma_j} \qquad\mbox{with}\quad \sigma_i\in\{1,2,...,q\} \, ,
\ee
which is symmetric under the group $G=S_q$ composed of the $\vert G\vert=q!$ permutations \cite{W82}.
The order parameter is here defined as
\be
{\bf M} = \{ M_s\}_{s=1}^q \qquad\mbox{with}\quad M_s = \sum_i \delta_{\sigma_i,s}
\ee
and the external field such that ${\bf B}\cdot{\bf M}=\sum_s B_s M_s = \sum_i B_{\sigma_i}$.

Monte Carlo simulations illustrate the application of the fluctuation relation (\ref{FR}) to the particular case $q=3$ of the Potts model of Hamiltonian
\be
H_{0}= - J \, \sum_{<i,j>} \delta_{\sigma_i,\sigma_j} \qquad\mbox{with}\quad \sigma_i\in\{0,1,2\}
\label{H0-3Potts}
\ee
with nearest-neighbor interaction on a square lattice with $N=L\times L$ sites ($i,j=1,2,...,N$).
The order parameter is taken as the magnetization:
\be
{\bf M} = \sum_i \left(\cos\frac{2\pi\sigma_i}{3},\sin\frac{2\pi\sigma_i}{3}\right) \, .
\ee
The Hamiltonian (\ref{H0-3Potts}) is symmetric under the groups ${\mathbb Z}_3$ and $C_{\rm 3v}$ \cite{H89}.
This symmetry is broken by an external magnetic field $\bf B$, in which case the Hamiltonian takes the form~(\ref{Hamilt}).

The canonical distribution at the inverse temperature $\beta=(kT)^{-1}$ is simulated by the Monte Carlo algorithm, in which a spin $\sigma_i$ is taken with equal probability on 
the $L^2$ sites of the square lattice and flipped to a new value $\sigma_i'$ with the probability $P(\sigma_i')=\exp\left[-\beta\Delta E(\sigma_i')\right]/Z$ where $\Delta E(\sigma_i')=H_{\bf B}(\pmb{\sigma}')-
H_{\bf B}(\pmb{\sigma})$ is the energy difference due to the spin flip and $Z=\sum_{\sigma_i'=0,1,2}\exp\left[-\beta\Delta E(\sigma_i')\right]$.

\begin{figure}[h]
\centerline{\scalebox{0.5}{\includegraphics{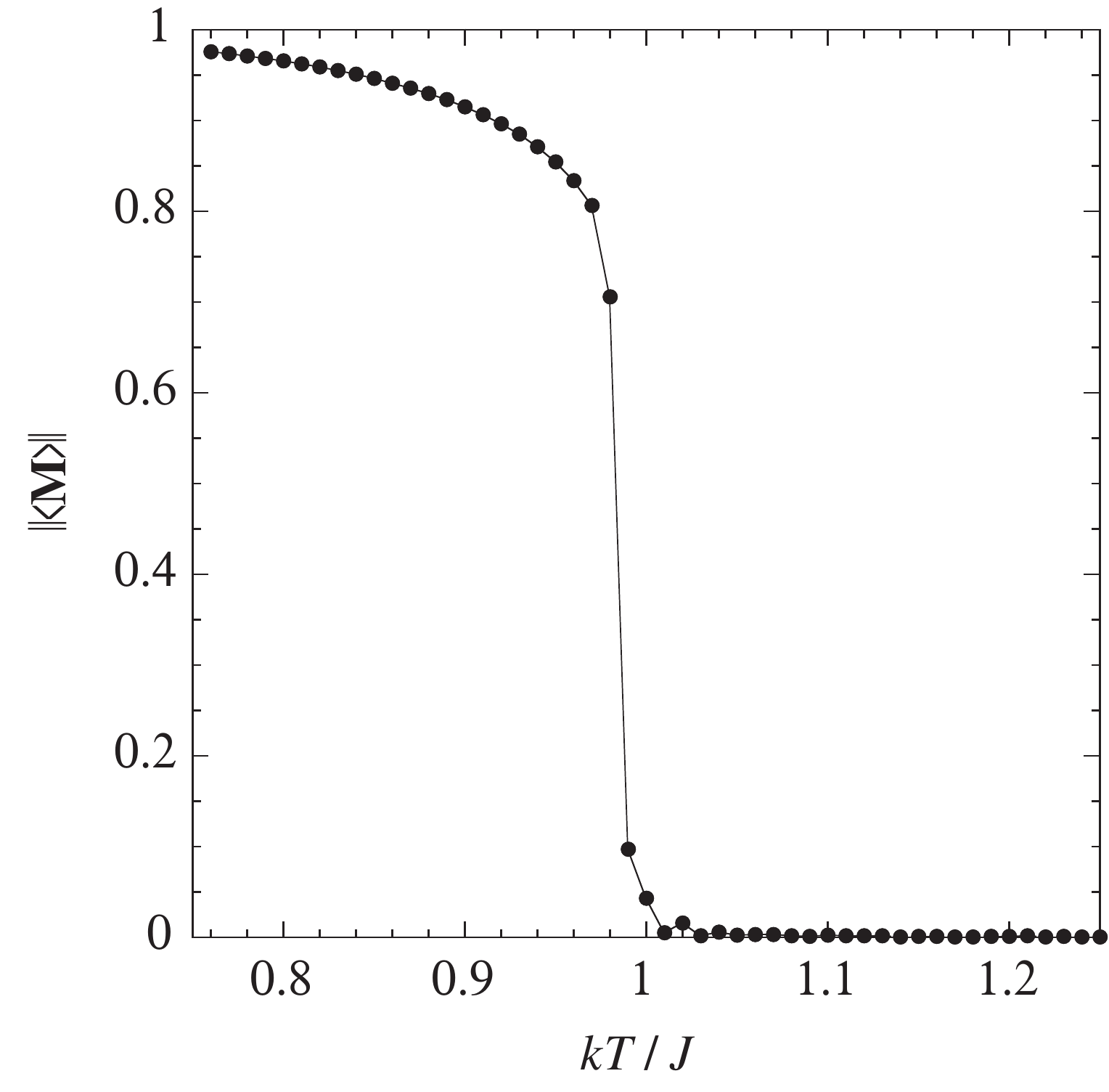}}}
\caption{$3$-state Potts model on a square lattice with $L=50$ in an external magnetic field $(B_x=10^{-4},B_y=0)$: magnitude of the average magnetization 
versus the rescaled temperature $kT/J$.  The statistical average is taken over $20000$ values of the magnetization.  
Every value is sampled after $125000$ spin flips.}
\label{fig7}
\end{figure}

Figure~\ref{fig7} depicts the magnitude of the average magnetization as a function of the rescaled temperature $kT/J$, showing that the paramagnetic-ferromagnetic phase transition happens at a temperature compatible with the known  critical value $kT_{\rm c}/J=1/\ln(1+\sqrt{3})=0.99497$ \cite{W82}.

\begin{figure}[h]
\centerline{\scalebox{0.5}{\includegraphics{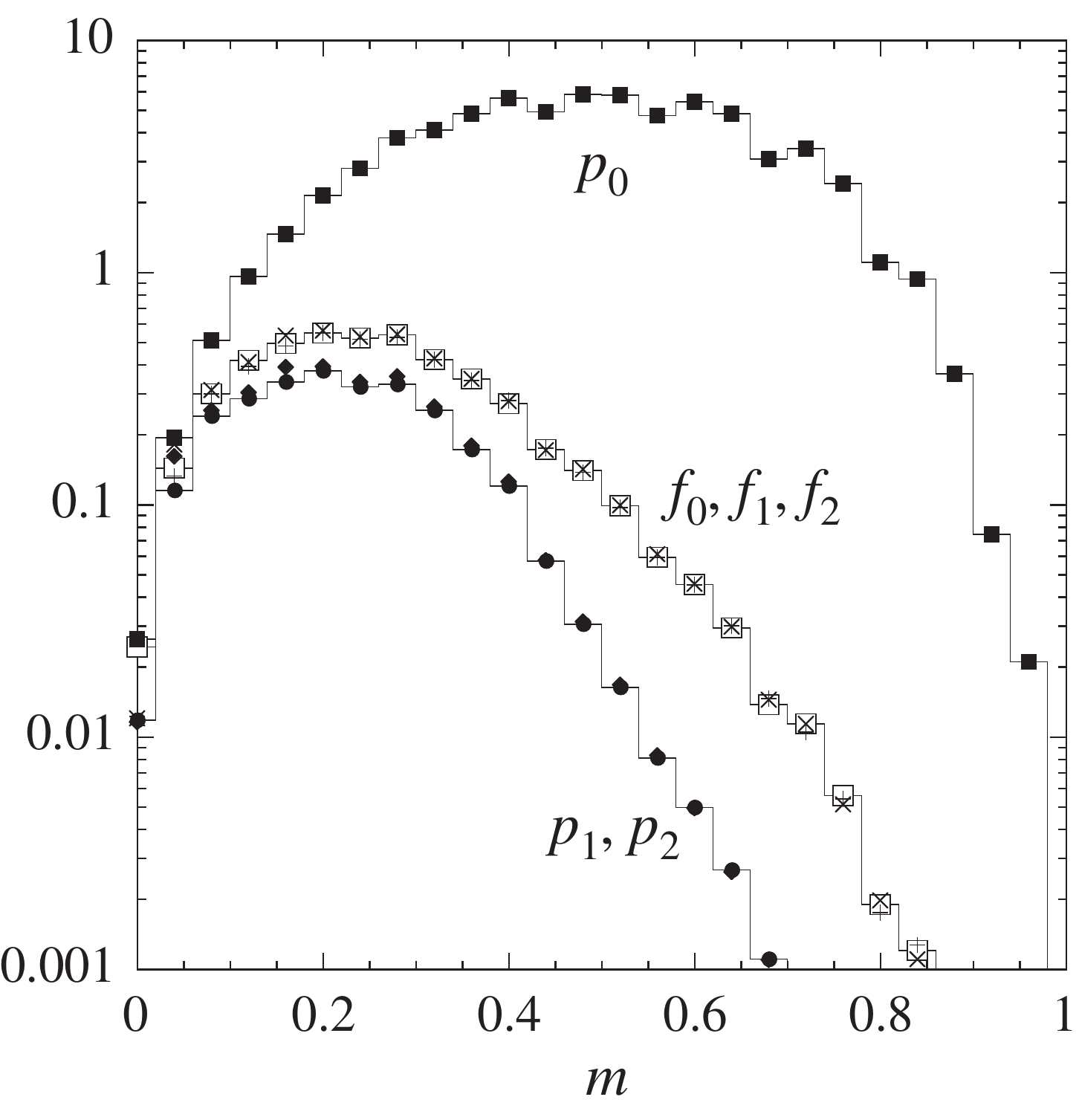}}}
\caption{$3$-state Potts model on a square lattice with $L=10$ in an external magnetic field $(B_x=0.1,B_y=0)$ at the rescaled temperature $kT/J=1.25$:  
The filled squares, the circles, and the diamonds depict the probability distributions (\ref{p_n}) of the magnetization per spin $m=M/L^2$ in the three angular
 sectors (\ref{3_sectors}), while the open squares, the crosses, and the pluses depict the distributions with the Boltzmann weights (\ref{f_n}).  
Their coincidence (\ref{f_1=f_2=f_3}) tests the fluctuation relation.  The statistical average is taken over $10^8$ values of the magnetization. 
 Every value is sampled after $1000$ spin flips.  The histograms are established with $\Delta m=1/25$.}
\label{fig8}
\end{figure}

In order to test the fluctuation relation, the probability distribution $P_{\bf B}({\bf M})$ is integrated over three angular sectors corresponding to the following values of the magnetization:
\be
{\bf M}= \left(M\cos\theta,M\sin\theta\right) \qquad\mbox{with} \quad \theta\in S_n=\left(-\frac{\pi}{3}+\frac{2\pi}{3}n,\frac{\pi}{3}+\frac{2\pi}{3}n\right) 
\label{3_sectors}
\ee
and $n=0,1,2$.  The probability distribution of the magnitude of the magnetization in the three sectors are thus defined as
\be
p_n(M)=\frac{3}{\Delta M} \int_{M<\Vert{\bf M}\Vert<M+\Delta M,\,\theta\in S_n} d{\bf M}\, P_{\bf B}({\bf M})
\label{p_n}
\ee
for $n=0,1,2$.  We also define the distributions with Boltzmann's weights as
\be
f_n(M)=\frac{3}{\Delta M} \int_{M<\Vert{\bf M}\Vert<M+\Delta M,\,\theta\in S_n} d{\bf M}\, P_{\bf B}({\bf M})\, {\rm e}^{-\beta{\bf B}\cdot{\bf M}}
\label{f_n}
\ee
for $n=0,1,2$.  The fluctuation relation (\ref{FR}) implies the equality of these three weighted distributions:
\be
f_1(M)=f_2(M)=f_3(M)\, .
\label{f_1=f_2=f_3}
\ee
This is tested in Fig.~\ref{fig8} plotting both the distributions (\ref{p_n}) and (\ref{f_n}).  The magnetic field points in the middle of the sector $n=0$ so that the distribution $p_0$ is larger than the two others.  Moreover, $p_1=p_2$ because the sectors $n=1$ and $n=2$ are located symmetrically with respect to the direction of the magnetic field.
Therefore, the three distributions~(\ref{p_n}) do not coincide in the presence of the external magnetic field.  In contrast, the three weighted distributions~(\ref{f_n}) coincide within numerical errors in agreement 
with the prediction of the fluctuation relation.

\subsection{$p$-state clock model}

Similar considerations apply to the closely related $p$-state clock model defined by the Hamiltonian
\be
H_{0}= - \sum_{i<j} J_{ij} \, \cos\frac{2\pi(n_i-n_j)}{p} \qquad\mbox{with}\quad n_i\in\{1,2,...,p\} \, .
\ee
This latter is symmetric under the group $G={\mathbb Z}_p=\{1,\eta,\eta^2,...,\eta^{p-1}\}$ generated by the transformation:
\be
\eta\, n_i = \left\{ \begin{array}{ll}
n_i+1 &\mbox{if}\ \ n_i\in\{1,2,...,p-1\} \, , \\
1 &\mbox{if}\ \ n_i=p \, ,
\end{array} \right.
\ee
which is a rotation by the angle $2\pi/p$.  For $p=3$, the clock model is equivalent to the $3$-state Potts model, which we have analyzed here above. The clock model interpolates between the Ising model ($p=2$) and the $XY$ model ($p=\infty$) \cite{W82}.

\subsection{The one-dimensional $p$-state clock chain}

Let us suppose that the spins or clocks form a one-dimensional chain.  Accordingly, the Hamiltonian is given by
\be
H_{0}= - J \sum_{i=1}^N \cos(\theta_i-\theta_{i+1}) \qquad\mbox{with}\quad \theta_i=\frac{2\pi n_i}{p} \qquad\mbox{and}\quad n_i\in\{0,1,2,...,p-1\} \, .
\label{H0-clock_chain}
\ee
The Hamiltonian $H_0$ is symmetric under the rotations by the angles $\alpha_n=2\pi n/p$ ($n=0,1,2,...,p-1$) such that
\be
{\bf M}'=\boldsymbol{\mathsf R}_n \cdot{\bf M} = \left(
\begin{array}{cc}
\cos\alpha_n & -\sin\alpha_n\\
\sin\alpha_n & \cos\alpha_n
\end{array}
\right)\cdot{\bf M}
\label{clock-rot}
\ee
We notice that the Hamiltonian $H_0$ is also symmetric under the reflections:
\be
{\bf M}'=\boldsymbol{\mathsf R}_{\bar{n}} \cdot{\bf M} = \left(
\begin{array}{cc}
\cos 2\alpha_n & \sin 2\alpha_n\\
\sin 2\alpha_n & -\cos 2\alpha_n
\end{array}
\right)\cdot{\bf M}
\label{clock-refl}
\ee
where $\alpha_n=2\pi n/p$ with $n=0,1,2,...,p-1$.  The full symmetry group is thus $G=C_{p{\rm v}}$ with $\vert G\vert=2p$ elements, which are $p$ rotations and $p$ reflections.

The coupling to the external magnetic field is again given by $H=H_0-\B\cdot\M$ with the same magnetization~(\ref{Magnet-XY}) as for the $XY$ model.  
The partition function can be expressed as
\be
Z_N({\bf B})={\rm tr}\, {\rm e}^{-\beta H} = {\rm tr} \left(\hat V _{{\bf B},{\bf 0}}\right)^N
\ee
in terms of a transfer operator ${\hat V}_{{\bf B},{\bf 0}}$.
On the other hand, the cumulant generating function~(\ref{GF}) is given in terms of the leading eigenvalue of another transfer operator defined as
\be
V _{{\bf B},\pmb{\lambda}}(n_i\vert n_{i+1}) = \exp\left[\beta J \cos(\theta_i-\theta_{i+1}) + (\beta B_x-\lambda_x)\frac{\cos\theta_i+\cos\theta_{i+1}}{2} + (\beta B_y-\lambda_y)\frac{\sin\theta_i+\sin\theta_{i+1}}{2}\right]
\ee
with $\theta_i=2\pi n_i/p$ and $\theta_{i+1}=2\pi n_{i+1}/p$.  

In the case of three states ($p=3$), the transfer operator takes the form of the following $3\times 3$ matrix:
\be
\hat V _{{\bf B},\pmb{\lambda}} = 
\left(
\begin{array}{ccc}
{\rm e}^{K+C_x} & {\rm e}^{-K/2+C_x/4+\sqrt{3} C_y/4} & {\rm e}^{-K/2+C_x/4-\sqrt{3} C_y/4} \\
{\rm e}^{-K/2+C_x/4+\sqrt{3} C_y/4} & {\rm e}^{K-C_x/2+\sqrt{3} C_y/2} & {\rm e}^{-K/2-C_x/2} \\
{\rm e}^{-K/2+C_x/4-\sqrt{3} C_y/4} & {\rm e}^{-K/2-C_x/2} & {\rm e}^{K-C_x/2-\sqrt{3} C_y/2}
\end{array}
\right)
\ee
with $K=\beta J$, $C_x=\beta B_x-\lambda_x$, and $C_y=\beta B_y-\lambda_y$.

The group elements of $C_{3{\rm v}}$ act on the three states also by $3\times 3$ matrices:
\bea
\hat R_0 = 
\left(
\begin{array}{ccc}
1 & 0 & 0 \\
0 & 1 & 0 \\
0 & 0 & 1
\end{array}
\right) \, ,
\qquad
\hat R_1 = 
\left(
\begin{array}{ccc}
0 & 1 & 0 \\
0 & 0 & 1 \\
1 & 0 & 0
\end{array}
\right) \, ,
\qquad
\hat R_2 = 
\left(
\begin{array}{ccc}
0 & 0 & 1 \\
1 & 0 & 0 \\
0 & 1 & 0
\end{array}
\right) \, ,
\nonumber\\
\hat R_{\bar{0}} = 
\left(
\begin{array}{ccc}
1 & 0 & 0 \\
0 & 0 & 1 \\
0 & 1 & 0
\end{array}
\right) \, ,
\qquad
\hat R_{\bar{1}} = 
\left(
\begin{array}{ccc}
0 & 0 & 1 \\
0 & 1 & 0 \\
1 & 0 & 0
\end{array}
\right) \, ,
\qquad
\hat R_{\bar{2}} = 
\left(
\begin{array}{ccc}
0 & 1 & 0 \\
1 & 0 & 0 \\
0 & 0 & 1
\end{array}
\right) \, .
\eea

A straightforward calculation shows that the transfer matrix also obeys the symmetry relation~(\ref{V-sym})
where $\boldsymbol{\mathsf R}_g$ are the $2\times 2$ matrices~(\ref{clock-rot})-(\ref{clock-refl}).
As a consequence, the cumulant generating function has the symmetry~(\ref{G-sym}) as required by the fluctuation relation.

\subsection{Anisotropic Curie-Weiss model}

Here, we consider $N$ classical spins with the mean-field Curie-Weiss interaction of Hamiltonian 
\be
H_N(\pmb{\sigma};{\bf B})=-\frac{1}{2N} \, {\bf M}_N(\pmb{\sigma})^{\rm T} \cdot \boldsymbol{\mathsf J} \cdot {\bf M}_N(\pmb{\sigma}) -{\bf B}^{\rm T} \cdot{\bf M}_N(\pmb{\sigma}) \, .
\ee
where $\boldsymbol{\mathsf J}$ is a $3\times 3$ matrix of coupling constants, which describes the anisotropy of the system. By 
construction, we only need to consider $\boldsymbol{\mathsf J}$ as a real symmetric matrix. After a change of basis, such a matrix can be diagonalized by an orthogonal transformation so that we may suppose without loss of generality that $\boldsymbol{\mathsf J}={\rm diag}(J_1,J_2,J_3)$ in the basis of its principal axes.

The relevant group which keeps the Hamiltonian $H_N(\pmb{\sigma};{\bf 0})$ invariant, is the group formed by elements whose representation satisfies
\be
\label{ani-Curie-group-rel}
\boldsymbol{\mathsf R}_g^{\rm T} \cdot \boldsymbol{\mathsf J} \cdot \boldsymbol{\mathsf R}_g = \boldsymbol{\mathsf J} \, .
\ee 
If all the three eigenvalues of that matrix are equal, the relevant group is clearly O(3) as in the case of the isotropic
Curie-Weiss studied before. Otherwise, the transformations satisfying Eq.~(\ref{ani-Curie-group-rel}) form a subgroup of O(3).

The distribution of the order parameter is rather similar to that of the isotropic Curie-Weiss model. It reads
\be
P_{\B}(\M)=\frac{1}{Z_N({\bf B})} \ {\rm e}^{\frac{\beta}{2N}\, {\bf M}^{\rm T} \cdot \boldsymbol{\mathsf J} \cdot {\bf M} +\beta{\bf B}^{\rm T} \cdot {\bf M}}  C_N({\bf M}) \, , 
\label{P(M)-ani}
\ee
with exactly the same function~(\ref{fn_C}) introduced before for the 
isotropic Curie-Weiss model. This is expected since this function is common to 
all mean-field models with this type of order parameter. Consequently, the entropy function is given by $S_N(N\m)=-k N I(m)$ 
with $I(m)$ defined in Eq.~(\ref{I(m)}). We recall that $m=\Vert\m\Vert$. One then obtains the large-deviation function as
\be
\label{Phi_B-ani}
\Phi_{\bf B}(\m) = I(m) - \frac{\beta}{2}\, \m^{\rm T} \cdot \boldsymbol{\mathsf J} \cdot \m - \beta\, \B^{\rm T} \cdot \m - \beta f(\B)\, .
\ee

If we assume that $\boldsymbol{\mathsf J}={\rm diag}(0,0,J_3)$, the relevant group is restricted to the subgroup $G={\rm O(2)}\otimes{\mathbb Z}_2$ of O(3). By minimizing $\Phi_{\bf B}(\m)$ with respect to $\m$, one obtains that $m_1=m_2=0$ and $m=m_3$ if $\B=(0,0,B)$.
The minimization with respect to the variable $m$ gives again the self-consistent equations $h_m = {\cal L}^{-1}(m) = I'(m)$ and $h_m=\beta J_3 m + \beta B$.  It follows from this, that the critical temperature is given by the same expression as obtained in the isotropic Curie-Weiss model, 
provided $J$ is replaced by $J_3$, in other words we have $kT_c=J_3/3$.  
The cumulant generating function $\Gamma_{\bf B}(\pmb{\lambda})$ can be obtained from $\Phi_{\bf B}(\m)$ by a Legendre-Fenchel transform.
When considering $\pmb{\lambda}=(0,0,\lambda)$ in the same direction as the magnetic field, the equations are 
exactly the same as that obtained for the isotropic Curie-Weiss model, 
provided $J$ is replaced by $J_3$. In particular, the behavior near criticality is the same.
The same exponent $\delta=3$ characterizes the non-analyticity of the cumulant generating function of both
the isotropic and anisotropic Curie-Weiss model, and by universality it characterizes all mean-field models possessing this
type of order parameter.

Using Eq.~(\ref{Phi_B-ani}), it is a simple matter to check that the fluctuation relation Eq.~(\ref{FR}) holds for this case, provided two values of the order parameter $\M$ and $\M'$ are related by
\be
\M'^{\rm T} \cdot \boldsymbol{\mathsf J} \cdot \M' = \M^{\rm T} \cdot \boldsymbol{\mathsf J} \cdot \M \, .
\ee
This relation is the exact equivalent of the relation derived in the non-equilibrium case for the fluctuations of the current in an anisotropic system 
\cite{VHT04}. As in the non-equilibrium case, the fluctuation relation corresponding to this anisotropic model is verified on ellipses 
in the order parameter space, as opposed to circles in the case of the isotropic Curie-Weiss model.

\section{Applications to nematic liquid crystals}
\label{Nematic}

\subsection{General derivation in the canonical ensemble}

Beyond magnetic systems, broken symmetry phases are ubiquitous in soft matter systems, in particular in liquid crystals, which are 
phases with broken rotational symmetry. 
These systems are of great interest to study deformations and orientation due to heterogeneities or to the application of external fields. 
Below, we focus on nematic liquid crystals which can be described by a tensorial order parameter ${\boldsymbol{\mathsf Q}}$ 
\cite{dGP93,Pi81,AnZa82}, or 
equivalently by a scalar order parameter and a director $\n$ for uniaxial nematics. 
As in the case of magnetic systems, we assume that a magnetic field is present, which breaks 
the symmetry that the Hamiltonian has in the absence of the field. 
The fact that this system has already a broken rotational symmetry when evaluated in the nematic phase 
even in the absence of the field,
does not affect the theoretical derivation of a fluctuation relation for the fluctuations of the order parameter, 
although it would greatly matter for the practical measurement of fluctuations. 
In view of this, future experimental tests of this result may
in fact be more easily done in the isotropic phase of the liquid crystal with a small magnetic field.
Alternatively, instead of considering a liquid crystal in 
the presence of external electric or magnetic fields,  
one could replace the magnetic field by an effective one associated with 
anchoring effects due to boundaries of the container.
We first discuss the fluctuations of the tensorial order parameter 
in a finite ensemble of nematogens and 
then we discuss the continuum description of long-wavelength 
distorsions of the director field $\n(\sr)$ in an extended system.

Let us consider the following general Hamiltonian:
\be
H_N(\pmb{\sigma};{\bf B}) = H_N(\pmb{\sigma};{\bf 0}) - {\bf B}^{\rm T}\cdot {\boldsymbol{\mathsf Q}}_N (\pmb{\sigma}) \cdot {\bf B},
\label{H-Q}
\ee
with the following traceless tensorial order parameter 
\be
\label{def-Q}
{\boldsymbol{\mathsf Q}}_N(\pmb{\sigma})=\sum_{i=1}^N 
\left(\pmb{\sigma}_i\otimes \pmb{\sigma}_i^{\rm T}-\frac{1}{d} \, {\boldsymbol{\mathsf 1}}\right)
\ee
where now $\pmb{\sigma}_i\in{\mathbb R}^d$ is a unit vector directed along the axis of the nematogens molecules. 
The distribution of this tensor is defined as
\be
P_{\bf B}({\boldsymbol{\mathsf Q}}) \equiv \langle \delta\left[{\boldsymbol{\mathsf Q}}-{\boldsymbol{\mathsf Q}}_N(\pmb{\sigma})\right]\rangle_{\bf B}
\label{P(Q)_nematics}
\ee
where $\langle\cdot\rangle_{\bf B}$ denotes the statistical average over Gibbs' canonical measure.
This probability density is normalized according to
\be
\int d{\boldsymbol{\mathsf Q}} \, P_{\bf B}({\boldsymbol{\mathsf Q}}) = 1 
\ee
where $d{\boldsymbol{\mathsf Q}}=\prod_{\mu,\nu=1,2,...,d} dQ_{\mu\nu}$.

Assuming that the Hamiltonian $H_N(\pmb{\sigma};{\bf 0})$ is symmetric under the transformations $g$ of a group $G$ in the absence of external field and using a similar derivation as before for a vectorial order parameter, one obtains the following isometric fluctuation relations for the distribution of the tensorial order parameter ${\boldsymbol{\mathsf Q}}$:
\be
P_{\bf B}({\boldsymbol{\mathsf Q}}) = P_{\bf B}({\boldsymbol{\mathsf Q'}}) \ {\rm e}^{\beta\, 
{\bf B}^{\rm T}\cdot({\boldsymbol{\mathsf Q}}-{\boldsymbol{\mathsf Q'}})\cdot{\bf B}}
\label{FT-Q-1}
\ee
with ${\boldsymbol{\mathsf Q'}}={\boldsymbol{\mathsf R}}_g^{-1}\cdot {\boldsymbol{\mathsf Q}}\cdot{\boldsymbol{\mathsf R}}_g^{-1{\rm T}}$ 
and
\be
P_{\bf B}({\boldsymbol{\mathsf Q}}) = P_{\bf B'}({\boldsymbol{\mathsf Q}}) \ {\rm e}^{\beta\, ({\bf B}^{\rm T}\cdot{\boldsymbol{\mathsf Q}}\cdot{\bf B}-{\bf B'}^{\rm T}\cdot{\boldsymbol{\mathsf Q}}\cdot{\bf B'})}
\label{FT-Q-2}
\ee
with ${\bf B'}={\boldsymbol{\mathsf R}}_g^{-1{\rm T}}\cdot {\bf B}$ for all $g\in G$.

It is interesting to note that the fluctuation relations of Eqs.~(\ref{FT-Q-1})-(\ref{FT-Q-2}) hold despite the fact that the symmetry breaking field $\B$ enters 
in non-linear way (here quadratically) in the Hamiltonian of Eq.~(\ref{H-Q}). This elementary but important observation shows that the fluctuation 
relations discussed in this paper are not limited to symmetry breaking fields entering linearly in the Hamiltonian. From the derivation of the fluctuation 
relations provided in this paper, it should be clear that, in fact, the symmetry breaking field can enter in an arbitrary non-linear way in the Hamiltonian
 and that the precise form of fluctuation relations will vary accordingly.

We also note that the isometric fluctuation relation~(\ref{FT-Q-1}) implies the inequality
\be
{\bf B}^{\rm T}\cdot\langle{\boldsymbol{\mathsf Q}}\rangle_{\bf B}\cdot{\bf B} \geq {\bf B}^{\rm T}\cdot{\boldsymbol{\mathsf R}}_g^{-1}\cdot\langle{\boldsymbol{\mathsf Q}}\rangle_{\bf B}\cdot{\boldsymbol{\mathsf R}}_g^{-1{\rm T}}\cdot{\bf B}\qquad\forall \ g\in G \, ,
\ee
which is deduced by using
\be
\int d{\boldsymbol{\mathsf Q}} \, P_{\bf B}({\boldsymbol{\mathsf Q}}) \, \ln \frac{P_{\bf B}({\boldsymbol{\mathsf Q}})}{P_{\bf B}({\boldsymbol{\mathsf Q'}})} \geq 0
\ee
with ${\boldsymbol{\mathsf Q'}}={\boldsymbol{\mathsf R}}_g^{-1}\cdot {\boldsymbol{\mathsf Q}}\cdot{\boldsymbol{\mathsf R}}_g^{-1{\rm T}}$.

\subsection{Maier-Saupe model of nematic liquid crystals}

To illustrate these results, we investigate a mean-field variant of the Maier-Saupe model \cite{MS58}, in which nematogens interact with a constant potential independent of their relative distance. Such a model has the Hamiltonian 
\be
H_N(\pmb{\sigma};{\bf B})=-\frac{J}{2N} \, {\rm tr}\, {\boldsymbol{\mathsf Q}}_N(\pmb{\sigma})^2 -{\bf B}^{\rm T}\cdot{\boldsymbol{\mathsf Q}}_N(\pmb{\sigma})\cdot{\bf B}\, .
\label{H-nematic}
\ee
In the absence of external field, the Hamiltonian $H_N(\pmb{\sigma};{\bf 0})$ is symmetric under the group O(3).  The probability distribution~(\ref{P(Q)_nematics}) of the order parameter is calculated using large-deviation theory in Appendix~\ref{AppA}.  

\begin{figure}[h]
\includegraphics[scale=0.45]{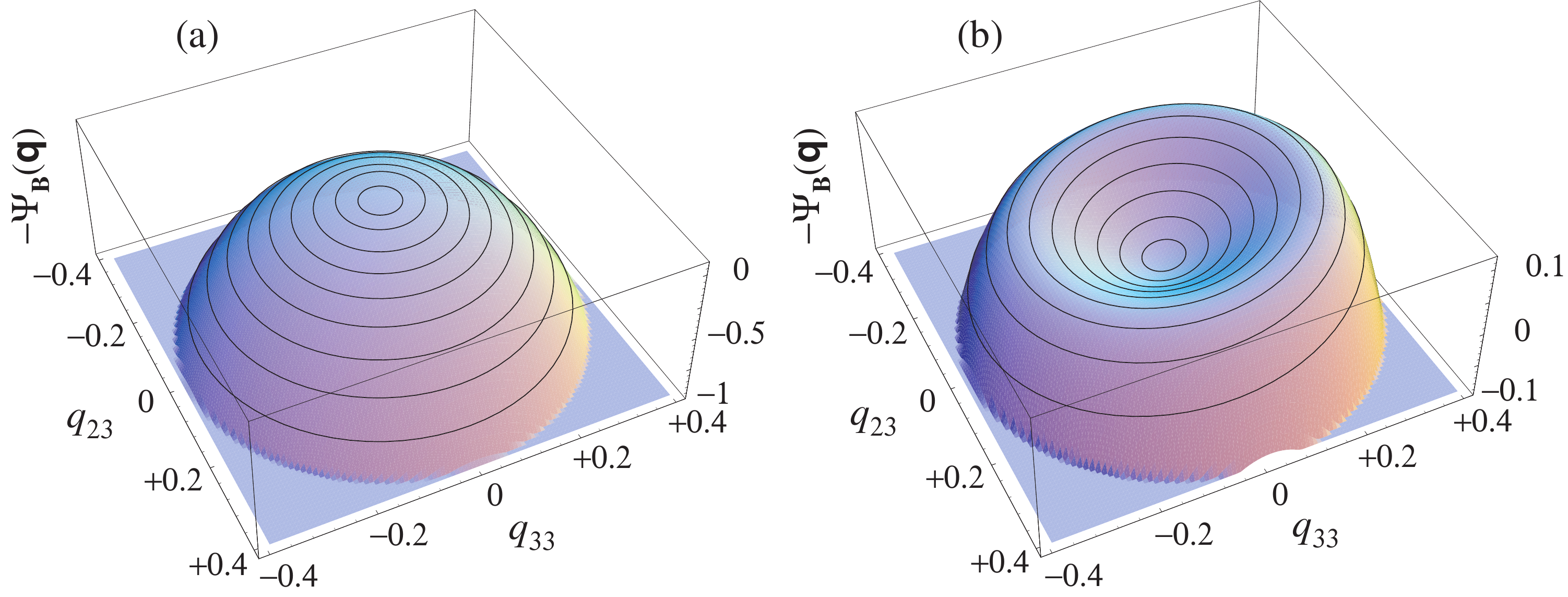}
\caption{Variant (\ref{H-nematic}) of the Maier-Saupe model of nematic liquid crystal with $J=1$ in the external field $\B=(0,0,0.1)$: Large-deviation function $-\Psi_\B({\boldsymbol{\mathsf q}})=-\Phi_\B({\boldsymbol{\mathsf q}})-\beta f(\B)$ with $\Phi_\B({\boldsymbol{\mathsf q}})=-\lim_{N\to\infty}(1/N)\ln P_\B(N{\boldsymbol{\mathsf q}})$ in the plane $q_{11}=q_{12}=q_{13}=0$ of the components $(q_{23},q_{33})$ at the rescaled inverse temperatures: (a) $\beta J=4$ in the isotropic phase and (b) $\beta J=10$ in the nematic phase.  The lines depict the contours of $(q_{23}^2+q_{33}^2)^{1/2}=0.04,0.08,...$ where the isometric fluctuation relation (\ref{FT-Q-1}) holds. See the detailed calculation in Appendix~\ref{AppA}.}
\label{fig9}
\end{figure}

In order to include biaxial fluctuations, which can be present even if the nematic liquid crystal 
is uniaxial on average, we parametrize the tensorial order parameter per nematogen ${\boldsymbol{\mathsf q}}\equiv{\boldsymbol{\mathsf Q}}/N$ as
\be
{\boldsymbol{\mathsf q}} = \frac{3}{2} \, s \, \left({\bf n}\otimes{\bf n}^{\rm T} - \frac{1}{3}\, {\boldsymbol{\mathsf 1}}\right) + \frac{1}{2} \, t \, \left({\bf l}\otimes{\bf l}^{\rm T} - {\bf m}\otimes{\bf m}^{\rm T}\right) ,
\label{def ST}
\ee
where the parameter $s$ represents the scalar order parameter of the nematic phase, $t$ measures 
the degree of biaxiality and $({\bf l},{\bf m},{\bf n})$ forms an orthonormal basis of unit vectors \cite{BMA10}. 
Figure~\ref{fig9} depicts the large-deviation function $\Phi_\B({\boldsymbol{\mathsf q}})=-\lim_{N\to\infty}(1/N)\ln P_\B(N{\boldsymbol{\mathsf q}})$ up to a constant, in the plane of two components of ${\boldsymbol{\mathsf q}}$ for the isotropic and nematic phases.  The isometric fluctuation relation (\ref{FT-Q-1}) is satisfied because the contour lines at equal values of $s=\sqrt{q_{23}^2+q_{33}^2}$ (which is also equal to $t$ in this case) coincide with the surface.  (See Appendix~\ref{AppA} for detail.)

\subsection{Fluctuation relation for a local tensorial order parameter}

In order to derive a fluctuation relation for the local tensorial order parameter, one proceeds exactly 
as for the case of a vectorial order parameter by coarse graining the order parameter density field defined by
\be
{\boldsymbol{\mathsf q}}(\sr)=\sum_{i=1}^N \left(\pmb{\sigma}_i\otimes \pmb{\sigma}_i^{\rm T}-\frac{1}{d} \, {\boldsymbol{\mathsf 1}}\right) 
\delta(\sr - \sr_i),
\ee 
where, as in the previous section, 
$\pmb{\sigma}_i\in{\mathbb R}^d$ is a unit vector directed along the axis of the nematogens molecules, and $\sr_i$ is the location of their center of mass. The volume $V$ is partitioned into small cells $\{\Delta V_j\}_{j=1}^c$ where this order parameter density ${\boldsymbol{\mathsf q}}(\sr)$ is coarse grained as
\be
{\boldsymbol{\mathsf q}}_j=\frac{1}{\Delta V_j}\int_{\Delta V_j} d\sr \, {\boldsymbol{\mathsf q}}(\sr) .
\ee
Moreover, the external magnetic field $\B(\sr)$ is supposed to be piecewise constant in the cells: 
$\B(\sr)=\B_j$ for $\sr\in\Delta V_j$. The joint probability distribution of the 
order parameter per nematogen in the cells is thus introduced as
\be
P_\B\left(\{ {\boldsymbol{\mathsf q}}_j \}\right) \equiv \left\langle\prod_{j=1}^{c} \delta\left[ {\boldsymbol{\mathsf q}}_j-\frac{1}{\Delta V_j} \int_{\Delta V_j} d\sr \, {\boldsymbol{\mathsf q}}(\sr)\right]\right\rangle_\B \, ,
\ee
where $\langle\cdot\rangle_\B$ is the statistical average over Gibbs' canonical probability distribution of Hamiltonian~(\ref{H-Q}). The interaction with the external field can be written as
\be
H_{\rm ext} = - \int_V d\sr \ \B(\sr)^{\rm T} \cdot {\boldsymbol{\mathsf q}}(\sr) \cdot \B(\sr) = 
-\sum_{j=1}^c \int_{\Delta V_j} d\sr \ \B(\sr)^{\rm T} \cdot {\boldsymbol{\mathsf q}}(\sr) \cdot \B(\sr) 
= -\sum_{j=1}^c \B_j^{\rm T} \cdot {\boldsymbol{\mathsf q}}_j \cdot \B_j \, \Delta V_j \, ,
\ee
so that the joint probability distribution takes the following form:
\be
P_\B\left(\{ {\boldsymbol{\mathsf q}}_j\}\right) = \frac{Z_N({\bf 0})}{Z_N({\bf B})}\; {\rm e}^{\beta\sum_{j=1}^c \B_j^{\rm T} \cdot {\boldsymbol{\mathsf q}}_j \cdot \B_j 
\, \Delta V_j} \; P_{\bf 0}\left(\{{\boldsymbol{\mathsf q}}_j\}\right) .
\ee
Since the Hamiltonian $H_N(\pmb{\sigma};{\bf 0})$ is symmetric under the group $G$, we obtain the fluctuation relation
\be
P_\B\left(\{ {\boldsymbol{\mathsf q}}_j\}\right) = P_\B\left(\{{\boldsymbol{\mathsf q'}}_j\}\right) \, {\rm e}^{\beta\sum_{j=1}^c \B_j^{\rm T} \cdot ({\boldsymbol{\mathsf q}}_j-{\boldsymbol{\mathsf q'}}_j) \cdot \B_j \, \Delta V_j} \, ,
\ee
where ${\boldsymbol{\mathsf q'}}_j={\boldsymbol{\mathsf R}}_g^{-1}\cdot {\boldsymbol{\mathsf q}}_j \cdot{\boldsymbol{\mathsf R}}_g^{-1{\rm T}}$.
In the limit where the cells of the partition are arbitrarily small, the joint probability distribution 
becomes the probability functional of the order parameter density and the sum turns into an integral so that 
the fluctuation relation for the local tensorial order parameter reads
\be
P_\B[ {\boldsymbol{\mathsf q}}(\sr)] = P_\B[{\boldsymbol{\mathsf q'}}(\sr)] \, {\rm e}^{\beta\int_V d\sr\, \B(\sr)^{\rm T} \cdot [ {\boldsymbol{\mathsf q}}(\sr)-{\boldsymbol{\mathsf q'}}(\sr)] \cdot \B(\sr)} \, ,
\ee
with ${\boldsymbol{\mathsf q'}}(\sr)={\boldsymbol{\mathsf R}}_g^{-1} \cdot {\boldsymbol{\mathsf q}}(\sr) \cdot{\boldsymbol{\mathsf R}}_g^{-1{\rm T}}$ for $g\in G$, as expected.

\subsection{Isometric fluctuation relation in the Frank-Oseen approach}

On larger length scales, the liquid crystal is described in terms of the field of its local director defined by the unit vector $\n({\bf r})$.  This field is supposed to extend over a spherical container of finite volume $V$ and to vary over scales larger than the mean distance between the nematogens. The free energy of the liquid crystal is modeled using the Frank-Oseen Hamiltonian \cite{CL95,GECL69}
\be
H[\n({\bf r});\B] = \frac{1}{2} \int_V d^3r \left\{ K_1 (\nabla \cdot \n)^2 + K_2 [\n \cdot (\nabla \times \n)]^2 + K_3 [\n \times (\nabla \times \n)]^2-\chi (\B \cdot \n)^2 \right\},
\label{Frank H}
\ee
where the parameters $K_i$ are three independent elastic constants of the liquid crystal 
and $\chi$ is the magnetic susceptibility. Note that the last term involves $(\B \cdot \n)^2$ instead of $\B \cdot \n$ due to the symmetry 
$\n \rightarrow -\n$ of nematic liquid crystals.

The fluctuation relations in which we are interested naturally hold for any value of the elastic constants.  First, we notice that the coupling 
to the external field in the Frank-Oseen Hamiltonian can be written as in Eq.~(\ref{H-Q}) 
if we introduce the tensorial order parameter
\be
{\boldsymbol{\mathsf Q}}=\int_V d^3r \left[\n({\bf r})\otimes \n({\bf r})^{\rm T}-\frac{1}{3} \, {\boldsymbol{\mathsf 1}}\right] .
\ee
As a consequence, the isometric fluctuation relations (\ref{FT-Q-1})-(\ref{FT-Q-2}) are satisfied for the probability distribution $P_{\bf B}({\boldsymbol{\mathsf Q}})$ in the Gibbsian canonical equilibrium state.  

Interestingly, a further relation can be obtained for the Fourier modes of the director fluctuations.  Considering a complete basis set of orthonormal functions $\{\varphi_{\bf k}({\bf r})\}$ in the spherical volume $V$,  the fluctuating field can be expanded as $\n({\bf r})=\sum_{\bf k} \tilde\n_{\bf k} \varphi_{\bf k}({\bf r})$ in terms of the Fourier amplitudes $\tilde\n_{\bf k}=\int_Vd^3r\,\varphi_{\bf k}^*({\bf r})\,\n({\bf r})$.  Consequently, the interaction with the external field $\bf B$ takes the quadratic form
\be
H_{\rm ext} = -\frac{\chi}{2} \int_V d^3r\, (\B \cdot \n)^2 = -\frac{\chi}{2} \sum_{\bf k}\vert\B\cdot\tilde\n_{\bf k}\vert^2 .
\ee
We define the joint probability distribution of the fluctuating amplitudes as $P_\B( \{ \pmb{\nu}_{\bf k} \} ) \equiv \langle \prod_{\bf k}\delta(\pmb{\nu}_{\bf k}-\tilde\n_{\bf k})\rangle_\B$ with respect to Gibbs' canonical measure for the Frank-Oseen Hamiltonian~(\ref{Frank H}).  Since this Hamiltonian is invariant under rotations ${\boldsymbol{\mathsf R}}_g$ in the absence of external field, a reasoning similar as above yields the following isometric fluctuation relation:
\be
P_\B ( \{ \pmb{\nu_{\rm k}} \})=P_\B (\{ \pmb{\nu_{\rm k}'} \}) \,{\rm e}^{\beta\chi\sum_{\bf k}(\vert\B\cdot\pmb{\nu_{\rm k}}\vert^2-\vert\B\cdot\pmb{\nu_{\rm k}'}\vert^2)/2}
\ee
with $\pmb{\nu_{\rm k}'}={\boldsymbol{\mathsf R}}_g^{-1}\cdot\pmb{\nu_{\rm k}}$.
A similar relation holds in particular for the reduced probability distribution of a single Fourier mode.
Note that this fluctuation relation could be potentially very useful since it concerns the individual Fourier modes $\pmb{\nu_{\rm k}}$ of the director in the liquid crystal,  and is compatible with the fact that the states $\{\pmb{\nu_{\rm k}}\}$ and $\{-\pmb{\nu_{\rm k}}\}$ cannot be distinguished in nematic liquid crystals. Note also that the above Hamiltonian represents only the bulk contribution of the energy. To test the relation, one needs in practice to consider a rotation that transforms $\n$ into ${\bf n'}$ without affecting the boundaries of the system. If surface contributions are important, they should be treated as the symmetry breaking field besides $\B$.

\subsection{Extension to the grand canonical ensemble}

Between the relation for a fixed number of nematogens in a given volume and that considered for a local order
parameter, a relation is here derived for a variable number of particles in a fixed volume,  
corresponding to grand canonical conditions. Such an extension could be useful to analyze
fluctuations of the liquid crystal order parameter near phase transitions that are driven by 
changes in densities, as in the case of the Onsager nematic-isotropic transition for instance.

The molecules are assumed to be non-spherical rigid bodies.  The position and momenta of their center of mass are denoted $\sr_i$ and ${\bf p}_i$.  The orientation of their direction $\s_i$ is determined by the Eulerian angles $\pmb{\alpha}_i$ and the corresponding momenta are denoted $\pmb{\pi}_i$.  The phase-space variables of a mechanical system of $N$ nematogens are thus given by $\pmb{\Gamma}_N=\{\sr_i,{\bf p}_i,\pmb{\alpha}_i,\pmb{\pi}_i\}_{i=1}^N$.  The order parameter of this system has been already defined in Eq.~(\ref{def-Q}) and the Hamiltonian of an ensemble of $N$ nematogens is assumed to be of the form $H_N(\pmb{\Gamma}_N;\B)=H_N(\pmb{\Gamma}_N;{\bf 0}) -\B^{\rm T} \cdot {\boldsymbol{\mathsf Q}}_N (\pmb{\Gamma}_N) \cdot \B$. 
The probability distribution of the order parameter now becomes a double sum over the number of particles and over the configurations with a fixed number of particles:
\ba
P_{\bf B}({\boldsymbol{\mathsf Q}}) &=& \frac{1}{\Xi({\bf B})} \sum_{N=0}^{\infty} {\rm e}^{\beta \mu N} \, \trace \,
{\rm e}^{-\beta H_N(\pmb{\Gamma}_N;{\bf 0})+\beta \B^{\rm T} \cdot {\boldsymbol{\mathsf Q}}_N (\pmb{\Gamma}_N) \cdot \B} \, \delta\left[{\boldsymbol{\mathsf Q}}-{\boldsymbol{\mathsf Q}}_N(\pmb{\Gamma}_N)\right], \nonumber\\
&=&\frac{\Xi({\bf 0})}{\Xi({\bf B})}\; {\rm e}^{\beta \B^{\rm T} \cdot {\boldsymbol{\mathsf Q}} \cdot \B} \; P_{\bf 0}({\bf M}),
\label{GI-LC}
\ea
where $\Xi(\B)$ is the grand canonical partition function and $\mu$ is the chemical potential. We have also introduced the trace over the configurations of the system of $N$ particles as 
\be
\trace (\cdot) = \frac{1}{N !} \int d\pmb{\Gamma}_N \, (\cdot) \, .
\ee

Now, we suppose that, in the absence of field, the Hamiltonian $H_N(\pmb{\Gamma}_N;{\bf 0})$ for a fixed number of nematogens is invariant under a symmetry group~$G$. 
As explained earlier, the group should act on all the phase-space variables of the nematogens, 
in other words, on their orientations, the positions of their center of mass, as well as the corresponding momenta.
Therefore,  $H_N(\pmb{\Gamma}_N^g;{\bf 0})= H_N(\pmb{\Gamma}_N;{\bf 0})$, where $\pmb{\Gamma}_N^g={\boldsymbol{\mathsf R}}_g\cdot\pmb{\Gamma}_N$, and ${\boldsymbol{\mathsf R}}_g$ is a representation of the element~$g$ of the group~$G$ such that $\vert\det{\boldsymbol{\mathsf R}}_g\vert=1$.  
As before, the probability distribution of the order parameter has this symmetry in the absence of magnetic field 
and
\ba
P_{\bf 0}({\boldsymbol{\mathsf Q}})&=& \frac{1}{\Xi({\bf 0})} \sum_{N=0}^{\infty} {\rm e}^{\beta \mu N} \, \trace\, {\rm e}^{-\beta H_N(\pmb{\Gamma}_N^g;{\bf 0})} \, \delta\left[{\boldsymbol{\mathsf Q}}-{\boldsymbol{\mathsf Q}}_N(\pmb{\Gamma}_N^g)\right], \nonumber\\
&=& \frac{1}{\Xi({\bf 0})} \sum_{N=0}^{\infty} {\rm e}^{\beta \mu N} \, \trace\, {\rm e}^{-\beta H_N(\pmb{\Gamma}_N;{\bf 0})} 
\, \delta\left[{\boldsymbol{\mathsf Q}}-{\boldsymbol{\mathsf R}}_g\cdot {\boldsymbol{\mathsf Q}}_N(\pmb{\Gamma}_N)
\cdot  {\boldsymbol{\mathsf R}}_g^{\rm T}   \right], \nonumber\\
&=& P_{\bf 0}\left({\boldsymbol{\mathsf R}}_g^{-1}\cdot {\boldsymbol{\mathsf Q}} \cdot {\boldsymbol{\mathsf R}}_g^{-1 {\rm T}} \right).
\label{proof-LC}
\ea
Combining Eqs.~(\ref{GI-LC}) and~(\ref{proof-LC}), one recovers the fluctuation relation of Eq.~(\ref{FT-Q-1}), which has now been derived for the grand canonical ensemble.

\section{Conclusion}
\label{Concl}

The present paper reports results in the continuation of our recent letter on isometric fluctuation relations for equilibrium systems \cite{LG14}.  
Here, these relations are applied to a broader selection of systems from equilibrium statistical mechanics, including not only the Curie-Weiss and $XY$~models 
of magnetism, and several models of nematic liquid crystals (for which we here give the detailed mathematical analysis), but also the one-dimensional 
Heisenberg chain, as well as anisotropic systems such as the $q$-state Potts model, the $p$-state clock model, and the anisotropic Curie-Weiss model.  
These latter systems show that the fluctuation relations apply not only to isotropic systems symmetric under a full rotational group O(2) or O(3), but
 also to anisotropic systems symmetric under a subgroup, which may be discrete or continuous.

This longer paper gives us the opportunity to present further developments and extensions of our previous results.  In particular, the relations 
are extended to general observables, which allows us to obtain the equilibrium analogues of the non-equilibrium Jarzynski and Crooks relations, and 
to show how the well-known fluctuation-response relations valid in the linear regime is recovered from the fluctuation relations.  In this perspective,
 these latter relations also contain information on the non-linear response properties of the system.  Besides, inequalities are deduced from the isometric
 fluctuation relations, which appear as the equilibrium analogues of Clausius' inequality.  Furthermore, the fluctuation relations are extended to equilibrium 
quantum systems.

An important issue is the understanding of the fluctuation relations as large-deviation properties in infinite equilibrium systems.  In this limit, 
large-deviation functions and their Legendre-Fenchel transforms can be introduced.  On the one hand, the well-known thermodynamic functions are recovered 
and, on the other hand, a class of large-deviation functions previously introduced by Landau and coworkers~\cite{LL60} can be defined systematically on the 
basis of probability theory.  It is in terms of these latter functions or their Legendre-Fenchel transforms that symmetry relations can be formulated in 
infinite equilibrium systems, starting from the isometric fluctuation relations obtained for finite equilibrium systems.

The possibility of spontaneous symmetry breaking turns out to result from the non-analyticity of the large-deviation function defining the cumulant generating
 function of the order parameter.  Its non-analyticity has universal scaling behavior determined by the critical exponents of equilibrium phase transitions.

Fluctuation relations can also be formulated for local order parameters, which are vectorial in magnetic systems, but tensorial in nematic ones.

For nematic liquid crystals, the calculation of large-deviation functions given in Appendix~\ref{AppA} uses methods from random matrix theory because 
the fluctuations of the tensorial order parameter are described by random matrices.  In this way, we obtain analytical expressions for the large-deviation
 functions describing the fluctuations of the tensorial order parameter in a mean-field variant of the Maier-Saupe model for uniaxial nematic liquid crystals.
  The analysis shows that the fluctuations may be biaxial, although the phase is uniaxial.

The equilibrium fluctuation relations can be deduced not only in the canonical ensemble, but also in the grand canonical ensemble, as we explicitly show 
for nematics.

Remarkably, the isometric fluctuation relations show that, even if broken, the fundamental symmetries continue to manifest themselves in the equilibrium 
fluctuations of the order parameters.  Accordingly, the fluctuation relations have many potential applications for the experimental investigation of the 
origins of broken symmetries from asymmetry measured in the fluctuations of some order parameter.  As we have here shown, the global or local fluctuations 
can be measured with this aim.  Experimental measurements based on fluctuation relations have already been performed in non-equilibrium systems to characterize 
the breaking of time-reversal symmetry \cite{AGCGJP08,TKVBMDLB14}.  In the light of the present results, we envisage that they can also be carried out at
 equilibrium.  The analysis of fluctuations in the critical regimes can be of upmost interest \cite{JPCG08}.

We have emphasized the analogies between the equilibrium and non-equilibrium fluctuation relations and other identities.  However, there 
also exist differences.  In particular, it is worthwhile to point out that, in finite equilibrium systems, the symmetry should be broken at the Hamiltonian
 level of description by introducing an external field.  Since Gibbs' canonical distribution is determined by the Hamiltonian, the symmetry is also broken 
at the statistical level of description.  For infinite systems, the symmetry remains broken in the absence of external field in the phenomenon of spontaneous
 symmetry breaking, in which case the symmetry is broken at the statistical level of description but no longer at the Hamiltonian level of description. 
 This situation is reminiscent of what happens in non-equilibrium systems where the Hamiltonian dynamics is microreversible and the time-reversal symmetry 
is only broken at the statistical level of description.

In conclusion, the fluctuation relations bring a unifying viewpoint on symmetry breaking beyond the traditional frontier between equilibrium and non-equilibrium
 systems.  Moreover, they provide a powerful method to detect hidden underlying symmetries among fluctuations.


\section*{Acknowledgments}

The authors thank P. Viot and K. Mallick for stimulating discussions, as well as 
D. Gaspard for the calculations of Eqs.~(\ref{I-2D_nem}) and (\ref{Jacobian-3D_nem}) with Mathematica \cite{W88}. 
P. Gaspard thanks the Belgian Federal Government for financial support under the Interuniversity Attraction Pole project P7/18~``DYGEST".


\appendix

\section{Mean-field models of nematics}
\label{AppA}

We consider the model of Hamiltonian (\ref{H-nematic}) with the extensive traceless tensor defined by Eq.~(\ref{def-Q}) in terms of the rotators $\pmb{\sigma}_i\in{\mathbb S}^{d-1}$ such that $\Vert\pmb{\sigma}_i\Vert=1$. The Hamiltonian $H_N(\pmb{\sigma};{\bf 0})$ is symmetric under the group O($d$).
In the space of the $N$ rotators, the invariant measure is defined with
\be
\int_{{\mathbb S}^{d-1}} \frac{d\pmb{\sigma}}{\Omega_d} \, (\cdot) \, .
\ee

The probability density that the tensor takes a given value ${\boldsymbol{\mathsf Q}}$ can be expressed as
\bea
P_{\bf B}({\boldsymbol{\mathsf Q}}) &\equiv& \langle \delta\left[{\boldsymbol{\mathsf Q}}-{\boldsymbol{\mathsf Q}}_N(\pmb{\sigma})\right]\rangle_{\bf B} \nonumber \\
&=& \frac{1}{Z_N({\bf B})} \, \int \frac{d^N\pmb{\sigma}}{\Omega_d^N} \, {\rm e}^{-\beta H_N(\pmb{\sigma};{\bf B})} \, \delta\left[{\boldsymbol{\mathsf Q}}-{\boldsymbol{\mathsf Q}}_N(\pmb{\sigma})\right] \nonumber \\
&=& \frac{1}{Z_N({\bf B})} \, {\rm e}^{\frac{\beta J}{2N}\, {\rm tr}\,{\boldsymbol{\mathsf Q}}^2 +\beta\,{\bf B}^{\rm T}\cdot{\boldsymbol{\mathsf Q}}\cdot{\bf B}} \, \int \frac{d^N\pmb{\sigma}}{\Omega_d^N} \, \delta\left[{\boldsymbol{\mathsf Q}}-{\boldsymbol{\mathsf Q}}_N(\pmb{\sigma})\right] \nonumber \\
&=& \frac{1}{Z_N({\bf B})} \ {\rm e}^{\frac{\beta J}{2N}\, {\rm tr}\,{\boldsymbol{\mathsf Q}}^2 +\beta\,{\bf B}^{\rm T}\cdot{\boldsymbol{\mathsf Q}}\cdot{\bf B}} \ C_N({\boldsymbol{\mathsf Q}})
\eea
and the partition function as
\bea
Z_N({\bf B}) &\equiv& \int \frac{d^N\pmb{\sigma}}{\Omega_d^N} \, {\rm e}^{-\beta H_N(\pmb{\sigma};{\bf B})} \nonumber \\
&=& \int d{\boldsymbol{\mathsf Q}}  \ {\rm e}^{\frac{\beta J}{2N}\, {\rm tr}\,{\boldsymbol{\mathsf Q}}^2 +\beta\,{\bf B}^{\rm T}\cdot{\boldsymbol{\mathsf Q}}\cdot{\bf B}} \ C_N({\boldsymbol{\mathsf Q}})
\eea
in terms of the function
\be
C_N({\boldsymbol{\mathsf Q}}) \equiv \int \frac{d^N\pmb{\sigma}}{\Omega_d^N} \ \delta\left[{\boldsymbol{\mathsf Q}}-{\boldsymbol{\mathsf Q}}_N(\pmb{\sigma})\right] ,
\label{C-Q-d}
\ee
which is normalized to unity according to
\be
\int d{\boldsymbol{\mathsf Q}} \, C_N({\boldsymbol{\mathsf Q}}) = 1 \, .
\ee

The function (\ref{C-Q-d}) can be calculated by using large-deviation theory.
The generating function of its statistical moments is introduced as
\be
\tilde C_N({\boldsymbol{\mathsf h}}) \equiv \int \frac{d^N\pmb{\sigma}}{\Omega_d^N} \ {\rm e}^{{\rm tr}\,{\boldsymbol{\mathsf h}}^{\rm T}\cdot{\boldsymbol{\mathsf Q}}_N(\pmb{\sigma})}
 = \int d{\boldsymbol{\mathsf Q}}\, {\rm e}^{{\rm tr}\,{\boldsymbol{\mathsf h}}^{\rm T}\cdot{\boldsymbol{\mathsf Q}}} \, C_N({\boldsymbol{\mathsf Q}})\, .
\label{C-h}
\ee
Since the tensor is defined by the sum~(\ref{def-Q}) over $N$ nematogens that are statistically independent according to the distribution~(\ref{C-Q-d}), the generating function is given by
\be
\tilde C_N({\boldsymbol{\mathsf h}}) =\chi({\boldsymbol{\mathsf h}})^N
\label{C-chi}
\ee
with
\be
\chi({\boldsymbol{\mathsf h}}) = {\rm e}^{-{\rm tr}\,{\boldsymbol{\mathsf h}}/d}
\int \frac{d\pmb{\sigma}}{\Omega_d} \ {\rm e}^{\pmb{\sigma}^{\rm T}\cdot {\boldsymbol{\mathsf h}}\cdot\pmb{\sigma}}\, .
\label{chi-Q-d}
\ee
Now, the idea is to express the function~(\ref{C-Q-d}) as
\be
C_N(N{\boldsymbol{\mathsf q}}) \sim {\rm e}^{-N\, I({\boldsymbol{\mathsf q}})}
\label{C-Q-d-LD}
\ee
in terms of the large-deviation function $I({\boldsymbol{\mathsf q}})$.  Inserting this assumption into Eq.~(\ref{C-h}) and combining with Eq.~(\ref{C-chi}), the large-deviation function is obtained as the following Legendre-Fenchel transform:
\be
I({\boldsymbol{\mathsf q}})={\rm Sup}_{\boldsymbol{\mathsf h}}\left[{\rm tr}\,{\boldsymbol{\mathsf h}}^{\rm T}\cdot{\boldsymbol{\mathsf q}}-\ln\chi({\boldsymbol{\mathsf h}})\right] .
\label{LF-gen}
\ee
A similar calculation can be performed using Fourier transforms.

For the following calculations, we point out that a general tensor ${\boldsymbol{\mathsf Q}}$ can be decomposed as
\be
{\boldsymbol{\mathsf Q}} = {\boldsymbol{\mathsf Q}}^{\rm S}+{\boldsymbol{\mathsf Q}}^{\rm A}
\label{Q-S-A}
\ee
into its symmetric and antisymmetric parts 
\bea
{\boldsymbol{\mathsf Q}}^{\rm S}&\equiv&\frac{1}{2} ({\boldsymbol{\mathsf Q}}+{\boldsymbol{\mathsf Q}}^{\rm T}) \, ,\\
{\boldsymbol{\mathsf Q}}^{\rm A}&\equiv&\frac{1}{2} ({\boldsymbol{\mathsf Q}}-{\boldsymbol{\mathsf Q}}^{\rm T}) \, .
\eea

\subsection{The two-dimensional case}

For $d=2$, the tensor ${\boldsymbol{\mathsf Q}}$ forms a $2\times 2$ matrix and its symmetric part can be diagonalized by an orthogonal transformation 
${\boldsymbol{\mathsf O}}$ (such as a rotation of angle $\theta$):
\be
{\boldsymbol{\mathsf Q}}^{\rm S} = {\boldsymbol{\mathsf O}}^{\rm T}\cdot
\left(
\begin{array}{cc}
\lambda_1 & 0 \\
0 & \lambda_2
\end{array}
\right)\cdot{\boldsymbol{\mathsf O}}
\ee
where
\be
(\lambda_1,\lambda_2) = \left( \frac{R}{2}+\frac{S}{2},\frac{R}{2}-\frac{S}{2}\right)
\ee
are the eigenvalues of the matrix ${\boldsymbol{\mathsf Q}}^{\rm S}$.  Since the order parameter is symmetric and traceless, $R=0$ and the 
Hamiltonian (\ref{H-nematic}) with the magnetic field $\B=(B,0)$ can be written in the following form:
\be
H_N(\s;\B) = -\frac{J}{4N} \, S^2 - \frac{B^2}{2} \, S \, \cos 2 \theta \, .
\ee

Besides, we notice that the integral in Eq.~(\ref{C-h}) is here carried out with the element of integration:
\be
d{\boldsymbol{\mathsf Q}}= 2\, d{\boldsymbol{\mathsf Q}}^{\rm S} \, d{\boldsymbol{\mathsf Q}}^{\rm A}
\ee
with
\be
d{\boldsymbol{\mathsf Q}}^{\rm S} =dQ_{11} \, dQ_{22} \, dQ_{12}^{\rm S} = \vert\lambda_1-\lambda_2\vert \, d\lambda_1 \, d\lambda_2 \, d\theta = \frac{1}{2} \, \vert S\vert \, dR \, dS \, d\theta \, ,
\label{Jacob1}
\ee
and $d{\boldsymbol{\mathsf Q}}^{\rm A}=dQ_{12}^{\rm A}$, which are obtained by calculating the Jacobian determinants of the changes of variables.

Now, with $\pmb{\sigma}=(\cos\phi,\sin\phi)$ and $\Omega_2=2\pi$, the integral (\ref{chi-Q-d}) becomes
\bea
\chi({\boldsymbol{\mathsf h}}) &=& {\rm e}^{-(h_{11}+h_{22})/2}
\int_0^{2\pi} \frac{d\phi}{2\pi} \ {\rm e}^{h_{11}\cos^2\phi+(h_{12}+h_{21})\cos\phi\sin\phi+h_{22}\sin^2\phi} \nonumber \\
&=& 
\int_0^{2\pi} \frac{d\phi}{2\pi} \ {\rm e}^{\frac{h_{11}-h_{22}}{2} \cos2\phi+\frac{h_{12}+h_{21}}{2}\sin2\phi} \nonumber \\
&=& I_0\left[\frac{1}{2}\sqrt{(h_{11}-h_{22})^2+(h_{12}+h_{21})^2}\right] ,
\eea
in terms of the modified Bessel function of zeroth order
\be
I_0(z) \equiv \frac{1}{\pi} \int_0^{\pi} d\alpha \, \exp(z\cos\alpha)\, .
\label{Bessel0}
\ee 
We notice that the function $\chi({\boldsymbol{\mathsf h}})$ does not depend on the trace and the asymmetric part of the matrix ${\boldsymbol{\mathsf h}}$, as expected.

Supposing that both matrices ${\boldsymbol{\mathsf h}}$ and ${\boldsymbol{\mathsf q}}$ are traceless symmetric, and denoting their respective eigenvalues by $\pm\varsigma/2$ and $\pm s/2$, we find that
\be
{\rm tr}\,{\boldsymbol{\mathsf h}}^{\rm T}\cdot{\boldsymbol{\mathsf q}} = \frac{1}{2} \, \varsigma \, s \, .
\ee
and the Legendre-Fenchel transform (\ref{LF-gen}) writes:
\be
I(s) = {\rm Sup}_{\varsigma} \left[ \frac{1}{2} \, \varsigma \, s - \ln I_0\left(\frac{\varsigma}{2}\right)\right] .
\ee
Expanding the functions in Taylor series, it is found that
\be
I(s) = s^2 + \frac{1}{4}\, s^4 + \frac{5}{36} \, s^6 + \frac{19}{192} \, s^8 + \frac{143}{1800} \, s^{10} + O(s^{12}) \, .
\label{I-2D_nem}
\ee
Accordingly, we get
\be
C_N(N{\boldsymbol{\mathsf q}}) \sim \delta(q_{11}+q_{22}) \, \delta(q_{12}-q_{21}) \, {\rm e}^{-N\, I\left(2\sqrt{q_{11}^2+q_{12}^2}\right)} .
\ee

\begin{figure}[h]
\includegraphics[scale=0.45]{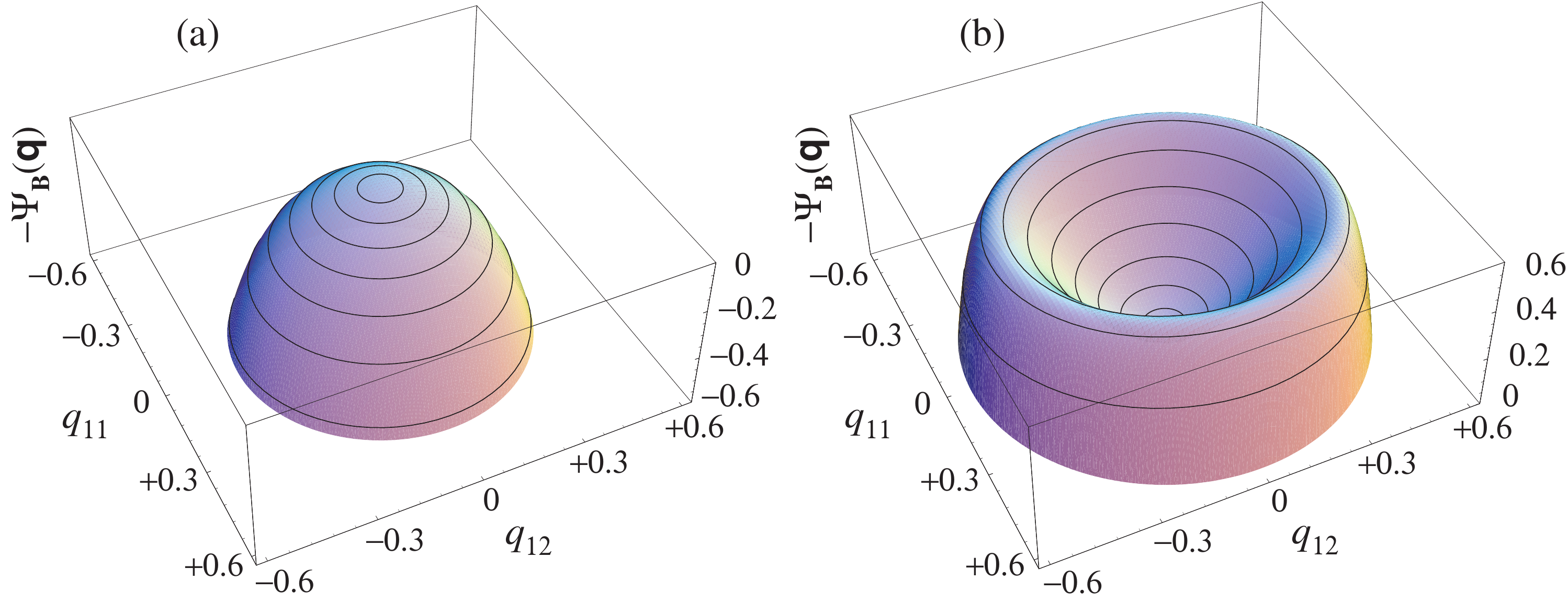}
\caption{Two-dimensional mean-field model of nematic liquid crystal with $J=1$ in the external field 
$\B=(0.1,0)$: Large-deviation function $-\Psi_\B({\boldsymbol{\mathsf q}})=-\Phi_\B({\boldsymbol{\mathsf q}})-\beta f(\B)$ with $\Phi_\B({\boldsymbol{\mathsf q}})=-\lim_{N\to\infty}(1/N)\ln P_\B(N{\boldsymbol{\mathsf q}})$ in the plane of the components $(q_{11},q_{12})$ at the rescaled inverse temperatures: (a) $\beta J=2$ in the isotropic phase and (b) $\beta J=8$ in the nematic phase.  The lines depict the contours of $(q_{11}^2+q_{12}^2)^{1/2}=0.06,0.12,...$ where the isometric fluctuation relation (\ref{FT-Q-1}) holds.}
\label{fig10}
\end{figure}

Therefore, the probability distribution of the tensorial order parameter at finite temperature and in the presence of an external field ${\bf B}=(B,0)$ is obtained as
\be
P_{\bf B}(N{\boldsymbol{\mathsf q}}) \sim {\rm e}^{-N \Phi_{\bf B}({\boldsymbol{\mathsf q}})}
\ee
with
\be
\Phi_{\bf B}({\boldsymbol{\mathsf q}}) = I\left(2\sqrt{q_{11}^2+q_{12}^2}\right) -\beta J\, (q_{11}^2+q_{12}^2)-\beta B^2 \, q_{11} -\beta f(\B) \, .
\ee
This function is depicted up to a constant in Fig.~\ref{fig10} together with isometric contour lines in order to show the validity of the fluctuation relation (\ref{FT-Q-1}) implying:
\be
\Phi_{\bf B}({\boldsymbol{\mathsf q}}) -\Phi_{\bf B}({\boldsymbol{\mathsf q'}}) =\beta B^2 (q'_{11}-q_{11})
\ee
where ${\boldsymbol{\mathsf q'}}$ is the tensor ${\boldsymbol{\mathsf q}}$ transformed by a rotation of O(2).  A key point is that these rotations preserve the expression $q_{11}^2+q_{12}^2=s^2/4$.  We notice that the phase transition between the isotropic to the nematic phases happens at the critical temperature $kT_c= J/4$ in this two-dimensional model.

\subsection{The three-dimensional case}

For $d=3$, the order parameter is a traceless symmetric tensor that can be written in the basis of its principal axes $({\bf l},{\bf m},{\bf n})$ after a rotation of SO(3) as follows \cite{BMA10}
\be
{\boldsymbol{\mathsf Q}} = \frac{3}{2} \, S \, \left({\bf n}\otimes{\bf n}^{\rm T} - \frac{1}{3}\, {\boldsymbol{\mathsf 1}}\right) + \frac{1}{2} \, T \, \left({\bf l}\otimes{\bf l}^{\rm T} - {\bf m}\otimes{\bf m}^{\rm T}\right) .
\label{tensor_Q}
\ee
The rotation diagonalizing the tensor can be expressed as ${\boldsymbol{\mathsf O}}={\boldsymbol{\mathsf O}}_\phi\cdot{\boldsymbol{\mathsf O}}_\theta\cdot{\boldsymbol{\mathsf O}}_\psi$ in terms of the Eulerian angles $(\phi,\theta,\psi)$ where ${\boldsymbol{\mathsf O}}_\phi$ and ${\boldsymbol{\mathsf O}}_\psi$ denote rotations around the $z$-axis by the angles $0\leq \phi,\psi \leq 2\pi$ and ${\boldsymbol{\mathsf O}}_\theta$ around the $y$-axis by the angle $0\leq \theta\leq \pi$.  In the basis of its principal axes $({\bf l},{\bf m},{\bf n})$, the tensor is diagonal and its eigenvalues are given by
\be
(\lambda_1,\lambda_2,\lambda_3) = \left( \frac{R}{3}-\frac{S}{2}+\frac{T}{2}, \frac{R}{3}-\frac{S}{2}-\frac{T}{2},\frac{R}{3} +S\right)
\label{eigenvalues}
\ee
where $R=0$ because the tensor is traceless.  The parameter $S$ is the scalar order parameter of the nematic phase.  This phase is uniaxial if $\langle T\rangle=0$, but the parameter $T$ may still have non-vanishing fluctuations.  In terms of these parameters, the Hamiltonian~(\ref{H-nematic}) is given by
\be
H_N = -\frac{J}{4N} (3S^2+T^2) - \frac{B^2}{2}\left[ S(3\cos^2\theta-1)+T\sin^2\theta\cos 2\phi\right] ,
\ee
since ${\bf B}\cdot{\bf l}=B\sin\theta\cos\phi$, ${\bf B}\cdot{\bf m}=B\sin\theta\sin\phi$, and ${\bf B}\cdot{\bf n}=B\cos\theta$.

For $d=3$, the integration element of the tensor is given by
\be
d{\boldsymbol{\mathsf Q}} = 8 \, d{\boldsymbol{\mathsf Q}}^{\rm S} \, d{\boldsymbol{\mathsf Q}}^{\rm A}
\ee
after decomposing the tensor into its symmetric and antisymmetric parts.  We find that $d{\boldsymbol{\mathsf Q}}^{\rm A}= dQ_{12}^{\rm A}\,dQ_{13}^{\rm A}\,dQ_{23}^{\rm A}$, while the integration element over the symmetric tensor can be 
written in terms of the eigenvalues~(\ref{eigenvalues}) and the Eulerian angles as
\ba
d{\boldsymbol{\mathsf Q}}^{\rm S} &=& dQ_{11}\,dQ_{22}\,dQ_{33}\,dQ_{12}^{\rm S}\,dQ_{13}^{\rm S}\,dQ_{23}^{\rm S} \nonumber\\
&=&\vert\lambda_1-\lambda_2\vert \, \vert\lambda_2-\lambda_3\vert \, \vert\lambda_3-\lambda_1\vert \, d\lambda_1 \, d\lambda_2 \, d\lambda_3 \, d\phi \, d\cos\theta \, d\psi \, ,
\label{Jacobian-3D_nem}
\ea
which is obtained by calculating the Jacobian determinant of the transformation 
${\boldsymbol{\mathsf Q}}^{\rm S}\to(\lambda_1,\lambda_2,\lambda_3,\phi,\theta,\psi)$ \cite{P65}.  The Eulerian angles are such
 that $\int_0^{2\pi}d\phi\int_0^{\pi}d\cos\theta\int_0^{2\pi}d\psi=8\pi^2$.  In terms of the parameters $(R,S,T)$, the integration element reads
\be
d{\boldsymbol{\mathsf Q}}^{\rm S} =\frac{1}{8} \vert T\vert \, \vert 3S+T\vert \, \vert 3S-T\vert \, dR \, dS \, dT \, d\phi \, d\cos\theta \, d\psi .
\ee

Now, the probability distribution~(\ref{C-Q-d}) is evaluated as above in the limit $N\to\infty$ by taking
\be
C_N(N{\boldsymbol{\mathsf q}}) = \delta(q_{12}-q_{21})\, \delta(q_{13}-q_{31})\, \delta(q_{23}-q_{32})\, \delta(q_{11}+q_{22}+q_{33}) \, D_N(s,t)
\label{C-delta}
\ee
with
\be
D_N(s,t) \simeq A_N(s,t) \, {\rm e}^{-NI(s,t)}
\label{D-Q}
\ee
where $s=S/N$, $t=T/N$, and the Dirac delta distributions are required to take into account the fact that the tensor ${\boldsymbol{\mathsf q}}={\boldsymbol{\mathsf Q}}/N$ is symmetric and traceless.

Here, the generating function (\ref{C-h}) is given by Eq.~(\ref{C-chi}) with
\be
\chi({\boldsymbol{\mathsf h}}) = {\rm e}^{-{\rm tr}\,{\boldsymbol{\mathsf h}}/3}
\int \frac{d\pmb{\sigma}}{4\pi} \ {\rm e}^{\pmb{\sigma}^{\rm T}\cdot {\boldsymbol{\mathsf h}}\cdot\pmb{\sigma}} .
\label{chi-Q-3}
\ee
where $d\pmb{\sigma}=d\cos\theta d\phi$ since $\pmb{\sigma}=(\sin\theta\cos\phi,\sin\theta\sin\phi\cos\theta)$ and $\Omega_3=4\pi$.

The complementary tensor ${\boldsymbol{\mathsf h}}$ can also be decomposed into its symmetric and antisymmetric parts as ${\boldsymbol{\mathsf h}}={\boldsymbol{\mathsf h}}^{\rm S}+{\boldsymbol{\mathsf h}}^{\rm A}$ with ${\boldsymbol{\mathsf h}}^{\rm S}=({\boldsymbol{\mathsf h}}+{\boldsymbol{\mathsf h}}^{\rm T})/2$ and ${\boldsymbol{\mathsf h}}^{\rm A}=({\boldsymbol{\mathsf h}}-{\boldsymbol{\mathsf h}}^{\rm T})/2$.  We deduce from Eq.~(\ref{chi-Q-3}) that $\chi({\boldsymbol{\mathsf h}})$ only depends on the symmetric part of~${\boldsymbol{\mathsf h}}$.  This symmetric tensor can be diagonalized as
\be
{\boldsymbol{\mathsf h}}^{\rm S} = {\boldsymbol{\mathsf O}}^{\rm T}\cdot{\boldsymbol{\eta}}\cdot{\boldsymbol{\mathsf O}}
\ee
in terms of the diagonal tensor ${\boldsymbol{\eta}}$ containing the eigenvalues
\be
(\eta_1,\eta_2,\eta_3) = \left( \frac{\rho}{3}-\frac{\varsigma}{2}+\frac{\tau}{2}, \frac{\rho}{3}-\frac{\varsigma}{2}-\frac{\tau}{2},\frac{\rho}{3} +\varsigma\right).
\ee
The calculation shows that the function~(\ref{chi-Q-3}) only depends on the complementary parameters $(\varsigma,\tau)$, but not on the trace ${\rm tr}\,{\boldsymbol{\mathsf h}}=\rho$:
\ba
\chi({\boldsymbol{\mathsf h}}) &=& \frac{1}{4\pi}\int d\cos\theta \, d\phi \, \exp\left[\frac{\varsigma}{2}(3\cos^2\theta-1)+\frac{\tau}{2}\sin^2\theta\cos 2\phi\right] \nonumber\\
&=& \int_0^1 d\xi \, {\rm e}^{\varsigma(3\xi^2-1)/2} \, I_0\left[\frac{\tau}{2}(1-\xi^2)\right] = \chi(\varsigma,\tau)
\ea
with the modified Bessel function of zeroth order (\ref{Bessel0}).

Using large-deviation theory and the G\"artner-Ellis theorem \cite{E85,E95,T09}, the rate function introduced in Eq.~(\ref{D-Q}) is given by the Legendre-Fenchel transform:
\be
I(s,t)={\rm Sup}_{\varsigma,\tau}\left[ \frac{3}{2}\, s \, \varsigma + \frac{1}{2}\, t \, \tau - \ln \chi(\varsigma,\tau)\right] .
\label{GE for LC}
\ee
Expanding in power series, we obtain
\be
I(s,t)= \frac{45}{8}\, s^2 +  \frac{15}{8}\, t^2 -  \frac{225}{56}\, s^3+ \frac{225}{56}\, s\, t^2 +\frac{34425}{3136}\, s^4 +\frac{14475}{1568}\, s^2 \, t^2 +\frac{3825}{3136}\, t^4 + \cdots
\label{I(s,t)}
\ee
neglecting terms of degree five or more.

As in the case of the Curie-Weiss model, the same result can be obtained via the variational mean-field approach. To do it in this way, we introduce the tensorial order parameter per nematogen ${\boldsymbol{\mathsf q}}\equiv{\boldsymbol{\mathsf Q}}/N$, and the average of this order parameter with respect to an {\it a priori} unknown probability distribution $\rho({\boldsymbol{\mathsf q}},\Omega)$ of the tensor ${\boldsymbol{\mathsf q}}$ in the solid angle~$\Omega$ as
\be
\langle {\boldsymbol{\mathsf q}} \rangle = \int d \Omega \, \boldsymbol{\mathsf q} \, \rho( {\boldsymbol{\mathsf q}}, \Omega) .
\label{SC}
\ee
The rotational entropy associated with the unknown distribution of the order parameter,
then enters in a mean-field variational free energy. By optimizing this variational mean-field free energy with respect to the unknown distribution, $\rho({\boldsymbol{\mathsf q}},\Omega)$, one finds
\be
\rho({\boldsymbol{\mathsf q}},\Omega) \sim \exp \left({\rm tr}\, {\boldsymbol{\mathsf h}}^{\rm T}\cdot {\boldsymbol{\mathsf q}} \right).
\ee
with the mean field
\be
 {\boldsymbol{\mathsf h}} = \beta J \langle{\boldsymbol{\mathsf q}}\rangle + \beta\, {\bf B}\otimes{\bf B}^{\rm T} .
\ee
One can then parametrize the average $\langle {\boldsymbol{\mathsf q}} \rangle$ from~(\ref{tensor_Q})  
and~${\boldsymbol{\mathsf q}}$ from Eq.~(\ref{def-Q}) in terms of the two scalar order parameters~$(s=S/N,t=T/N)$, and the Eulerian angles of the rotation diagonalizing $\langle {\boldsymbol{\mathsf q}} \rangle$ with respect to~${\boldsymbol{\mathsf q}}$.  A self-consistent equation is then obtained using Eq.~(\ref{SC}).
It is then straightforward to verify that the components of the self-consistent equation yield only two independent equations, which are identical with the two equations determining the extremum of Eq.~(\ref{GE for LC}). We thus have another example of the equivalence of the rate function of a large-deviation function at equilibrium with the rotational entropy function.

\begin{figure}[h]
\centerline{\scalebox{0.5}{\includegraphics{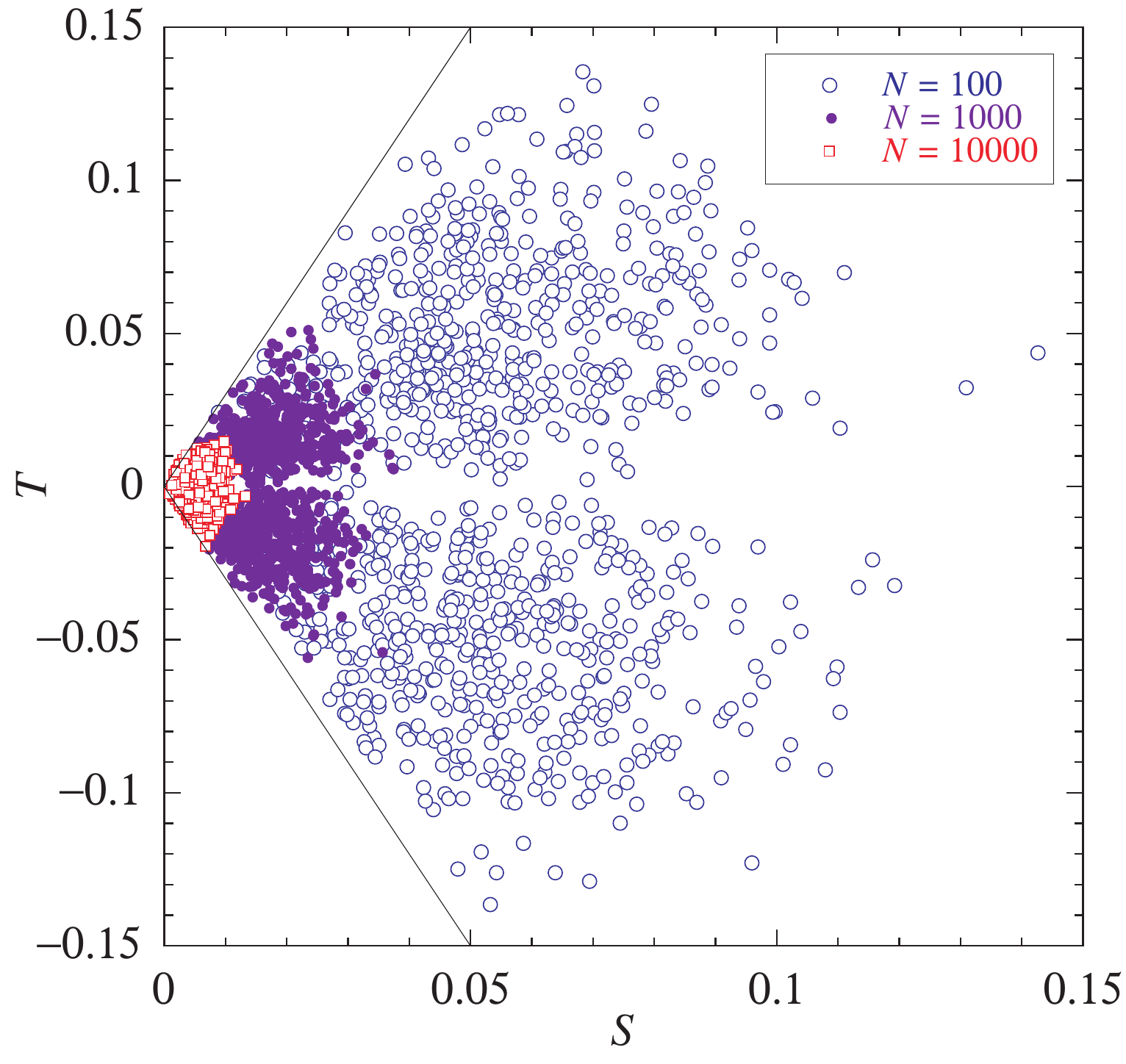}}}
\caption{The largest eigenvalue $S$ and the difference $T$ between the two smallest eigenvalues for a thousand of tensors ${\boldsymbol{\mathsf Q}}$ distributed according to $C_N({\boldsymbol{\mathsf Q}})$ in the plane $(S,T)$.  The points are limited to the domain $3S\ge T \ge -3S$.}
\label{fig11}
\end{figure}

Monte Carlo simulations have been performed in order to test the validity of the analytical results on the distribution $C_N({\boldsymbol{\mathsf Q}})$. Figure~\ref{fig11} shows the distribution of the order parameters $S$ and $T$ of $10^3$ traceless tensors ${\boldsymbol{\mathsf Q}}$ defined by Eq.~(\ref{def-Q}) for systems with $N=100$, $N=1000$, and $N=10000$ nematogens.  The order parameters are given by $S=\lambda_3$ and $T=\lambda_1-\lambda_2$ in terms of the eigenvalues (\ref{eigenvalues}).  Since they are ordered as $\lambda_3\geq \lambda_1,\lambda_2$, the inequalities $3S\ge T \ge -3S$ are satisfied.  We observe that both $S$ and $T$ are statistically distributed so that the fluctuations are biaxial although, on average, the system is uniaxial because the distribution of $T$ is symmetric under $T\to -T$ and thus of zero average.  We also observe that the fluctuations of both $S$ and $T$ goes as $1/\sqrt{N}$ for $N\to\infty$, as expected.

\begin{figure}[h]
\begin{center}
\includegraphics[scale=0.5]{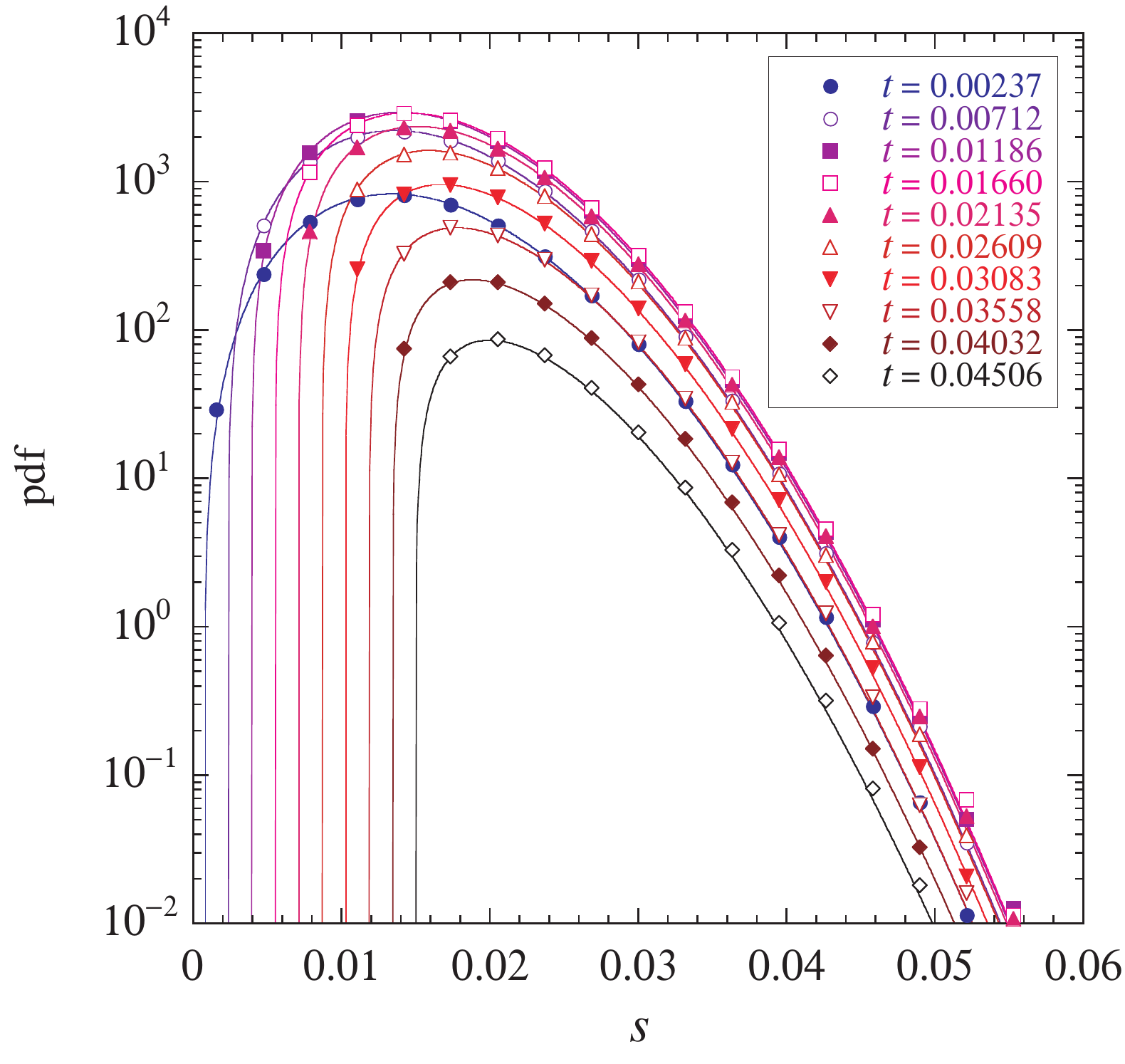}
\caption{Probability density~(\ref{D-asympt}) versus the parameters $s$ and $t$ obtained by Monte Carlo simulations (dots) compared with the theoretical calculation (lines) in terms of the rate function~(\ref{I(s,t)}).}
\label{fig12}
\end{center}
\end{figure}

Now, the distribution function~(\ref{D-Q}) of the parameters $(s,t)$ is given in the limit $N\to\infty$ by
\be
D_N(s,t) \simeq K \, \vert t\vert \, \vert 3s+t\vert \, \vert 3s-t\vert \, {\rm e}^{-N I(s,t)} 
\label{D-asympt}
\ee
where $K$ is a normalization constant such that $\int ds \, dt \, D_N(s,t) = 1$.
The support of this distribution is the domain with $s\geq 0$ and $\vert t\vert \leq 3s$.
This result has been tested by Monte Carlo simulations where $10^8$ random matrices~(\ref{tensor_Q}) with $N=1000$ have been generated and their eigenvalues calculated to obtain the histogram of the parameters $s=S/N$ and $t=T/N$.  In Fig.~\ref{fig12}, this histogram (dots) is compared with the theoretical result (lines) given by Eqs.~(\ref{D-asympt}) and~(\ref{I(s,t)}).  We observe the nice agreement validating the theoretical results.

Thanks to these results, we can now calculate the probability distribution of the tensorial order parameter at finite temperature and in the presence of an external field $\bf B$ in terms of $s=S/N$, $t=T/N$, and the Eulerian angles $(\phi,\theta,\psi)$ of the rotation diagonalizing the order parameter tensor in the basis $({\bf l},{\bf m},{\bf n})$ where
\bea
{\bf B}\cdot{\bf l} &=& B \cos\phi\sin\theta \, ,\\
{\bf B}\cdot{\bf m} &=& B \sin\phi\sin\theta \, ,\\
{\bf B}\cdot{\bf n} &=& B \cos\theta \, .
\eea
We have that
\bea
{\rm tr}\,{\boldsymbol{\mathsf q}}^2 &=& \frac{3}{2} \, s^2 + \frac{1}{2} \, t^2  \, ,\\
{\bf B}^{\rm T}\cdot{\boldsymbol{\mathsf q}}\cdot{\bf B} &=& \frac{1}{2} \, B^2 \left[ s (3\cos^2\theta-1) + t \sin^2\theta\cos2\phi\right] ,
\eea
so that the probability distribution is obtained as
\be
P_{\bf B}(N{\boldsymbol{\mathsf q}}) \sim {\rm e}^{-N \Phi_{\bf B}({\boldsymbol{\mathsf q}})}
\ee
with
\ba
\Phi_{\bf B}({\boldsymbol{\mathsf q}}) &=& I(s,t) - \frac{\beta J}{2}\, {\rm tr}\, {\boldsymbol{\mathsf q}}^2 -\beta {\bf B}^{\rm T}\cdot{\boldsymbol{\mathsf q}}\cdot{\bf B} -\beta f(\B) \nonumber\\
&=& I(s,t) - \frac{\beta J}{4}\, (3s^2+t^2) -\frac{\beta B^2}{2}\left[ s(3\cos^2\theta-1) + t \sin^2\theta \cos2\phi\right] -\beta f(\B) \, .
\ea

The isometric fluctuation relation can here be written in the form
\be
\Phi_{\bf B}({\boldsymbol{\mathsf q}}) -\Phi_{\bf B}({\boldsymbol{\mathsf q'}}) =\beta{\bf B}^{\rm T}\cdot({\boldsymbol{\mathsf q'}}-{\boldsymbol{\mathsf q}})\cdot{\bf B} \, .
\ee
This relation can be checked in the plane $q_{11}=q_{12}=q_{13}=0$ of the parameters $(q_{23},q_{33})$ with the external field ${\bf B}=(0,0,B)$.  We thus take
\be
{\boldsymbol{\mathsf q}} = 
\left(
\begin{array}{ccc}
0 & 0 & 0 \\
0 & -q_{33} & q_{23} \\
0 & q_{23} & q_{33}
\end{array}
\right)
\ee
and ${\boldsymbol{\mathsf q'}}={\boldsymbol{\mathsf O}}^{\rm T}\cdot{\boldsymbol{\mathsf q}}\cdot{\boldsymbol{\mathsf O}}$ with
\be
{\boldsymbol{\mathsf O}} = 
\left(
\begin{array}{ccc}
1 & 0 & 0 \\
0 & \cos\alpha & \sin\alpha \\
0 & -\sin\alpha & \cos\alpha
\end{array}
\right) .
\ee
We get
\be
\Phi_{\bf B}({\boldsymbol{\mathsf q}}) = I\left(\sqrt{q_{23}^2 + q_{33}^2},\sqrt{q_{23}^2 + q_{33}^2}\right) -\beta J (q_{23}^2 + q_{33}^2) - \beta B^2 q_{33} -\beta f(\B) \, .
\label{Phi-Q}
\ee
The parameters here take the values
\be
s=t=\sqrt{q_{23}^2 + q_{33}^2} \, .
\ee
This function is depicted up to a constant in Fig.~\ref{fig9} of the main text.  The isometric fluctuation relation is verified by plotting the contour lines where
\be
q_{23}^2 + q_{33}^2 = {q'_{23}}^{2} + {q'_{33}}^2 
\ee
and 
\be
\Phi_{\bf B}({\boldsymbol{\mathsf q}}) -\Phi_{\bf B}({\boldsymbol{\mathsf q'}}) =\beta B^2 ({q'_{33}}-q_{33})
\ee
should hold together.  This is the case because the contour lines in Fig.~\ref{fig9} of the main text indeed belong to the surface~(\ref{Phi-Q}).

\begin{figure}[h]
\centerline{\scalebox{0.45}{\includegraphics{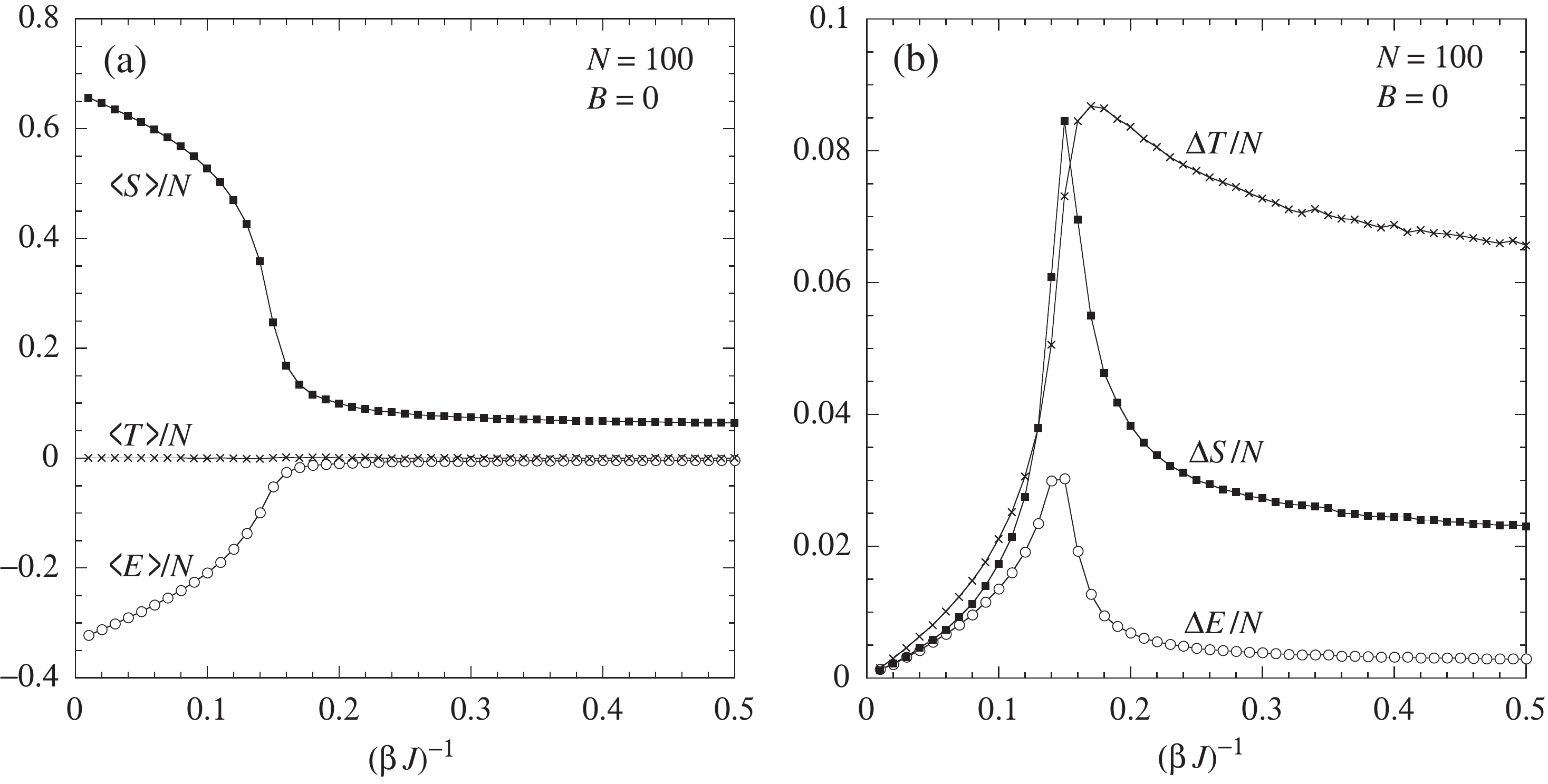}}}
\caption{3D system with $N=100$ nematogens and zero external field ${\bf B}=0$: (a) Average values and (b) root mean squares of the energy $E$ and the order parameters $S$ and $T$ versus the rescaled temperature $(\beta J)^{-1}$.}
\label{fig13}
\end{figure}

\begin{figure}[h]
\centerline{\scalebox{0.45}{\includegraphics{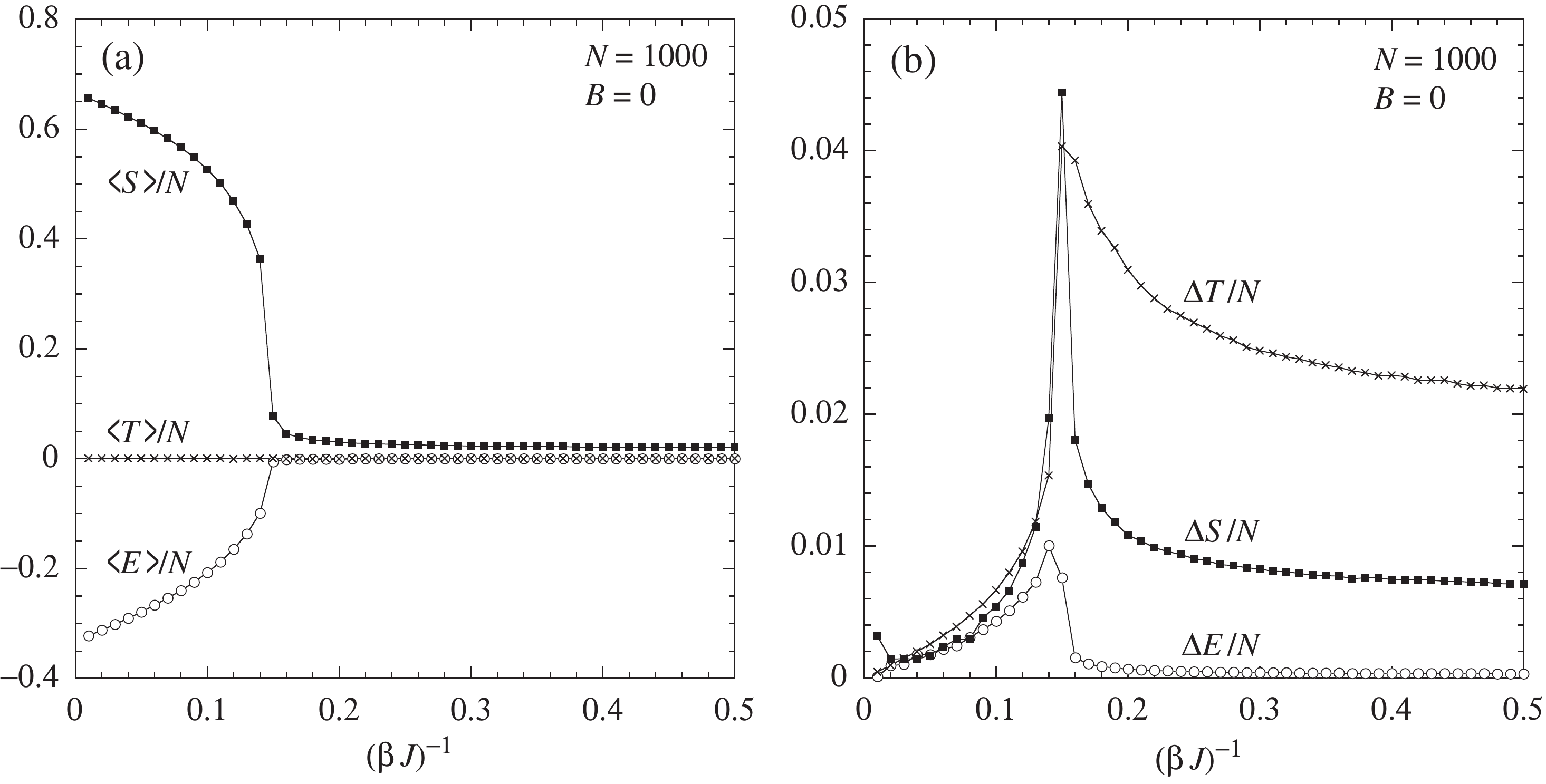}}}
\caption{3D system with $N=1000$ nematogens and zero external field ${\bf B}=0$: (a) Average values and (b) root mean squares of the energy $E$ and the order parameters $S$ and $T$ versus the rescaled temperature $(\beta J)^{-1}$.}
\label{fig14}
\end{figure}

We notice that this three-dimensional model corresponds to the Maier-Saupe model, which manifests a first-order phase transition from an isotropic phase to a uniaxial nematic phase beyond the critical value $\beta_{\rm c}J=6.8122$.  In the isotropic liquid phase, the nematic phase is metastable for $6.7315 \leq \beta J\leq \beta_{\rm c}J$.

The temperature dependence of the energy and order parameters has been simulated by the Metropolis algorithm.  
The statistics is performed over $2\times 10^4$ values for the different quantities of interest, namely, the energy $E$ and the order
 parameters $S$ and $T$.  Every set of values is sampled after $2\times 10^3$ random nematogen rotations.  The results are shown in Figs.~\ref{fig13}-\ref{fig14} where we observe that the system is indeed uniaxial on average because $\langle T\rangle=0$.  However, we see in Figs.~\ref{fig13}b-\ref{fig14}b that both order parameters $S$ and $T$ have non-vanishing fluctuations so that the fluctuations are biaxial.  The comparison between Fig.~\ref{fig13} and Fig.~\ref{fig14} shows that the root mean squares go as $1/\sqrt{N}$ for $N\to\infty$, as it should.



\begin{thebibliography}{99}

\bibitem{ECM93} D.~J.~Evans, E.~G.~D.~Cohen, and G.~P.~Morriss, Phys. Rev. Lett. {\bf 71}, 2401 (1993).

\bibitem{GC95} G.~Gallavotti and E.~G.~D.~Cohen, Phys. Rev. Lett. {\bf 74}, 2694 (1995).

\bibitem{K98} J.~Kurchan, J.~Phys.~A: Math. Gen. {\bf 31}, 3719 (1998).

\bibitem{C99} G.~E.~Crooks, Phys. Rev. E {\bf 60}, 2721 (1999).

\bibitem{LS99} J.~L.~Lebowitz and H.~Spohn, J. Stat. Phys. {\bf 95}, 333 (1999).

\bibitem{M99}  C.~Maes, J. Stat. Phys. {\bf 95}, 367 (1999).

\bibitem{AG06JSM} D.~Andrieux and P.~Gaspard, J. Stat. Mech.: Th. Exp. P01011 (2006).

\bibitem{AG07} D.~Andrieux and P.~Gaspard, J. Stat. Mech.: Th. Exp. P02006 (2007).

\bibitem{EHM09} M.~Esposito, U.~Harbola, and S.~Mukamel, Rev. Mod. Phys. {\bf 81}, 1665 (2009).

\bibitem{J11} C.~Jarzynski, Annu. Rev. Condens. Matter Phys. {\bf 2}, 329 (2011).

\bibitem{S12} U.~Seifert, Rep. Prog. Phys. {\bf 75}, 126001 (2012).

\bibitem{N09} Y. Nambu, Rev. Mod. Phys. {\bf 81}, 1015 (2009).

\bibitem{GSW62} J.~Goldstone, A.~Salam, and S.~Weinberg, Phys. Rev. {\bf 127}, 965 (1962).

\bibitem{A72} P.~W.~Anderson, Science {\bf 177}, 4047 (1972).

\bibitem{A84} P.~W.~Anderson, {\it Basic Notions of Condensed Matter Physics} (Benjamin/Cummings Publ. Co., Menlo Park CA, 1984).

\bibitem{F75} D. Forster, {\it Hydrodynamic Fluctuations, Broken Symmetry, and Correlation Functions} (Benjamin/Cummings Publ. Co., Reading MA, 1975).

\bibitem{CL95} P.~M.~Chaikin and T.~C.~Lubensky, {\it Principles of Condensed Matter Physics} (Cambridge University Press, Cambridge UK, 1995).

\bibitem{BE07} R. Brout and F. Englert, C. R. Physique {\bf 8}, 973 (2007).

\bibitem{H14} P. W. Higgs, Rev. Mod. Phys. {\bf 86}, 851 (2014).

\bibitem{PN67} I. Prigogine and G. Nicolis, J. Chem. Phys. {\bf 46}, 3542 (1967).

\bibitem{PLGH69} I. Prigogine, R. Lefever, A. Goldbeter, and M. Herschkowitz-Kaufman, Nature {\bf 223}, 913 (1969).

\bibitem{CH93} M. C. Cross and P. C. Hohenberg, Rev. Mod. Phys. {\bf 65}, 851 (1993).

\bibitem{PJ15} C. Presilla and G. Jona-Lasinio, Phys. Rev. A {\bf 91}, 022709 (2015).

\bibitem{GF92} N. Goldenfeld, {\it Lectures on Phase Transitions and the Renormalization Group}, (Perseus Books Publishing, L.L.C., 1992).

\bibitem{K10} J. Kurchan, {\it Six out of equilibrium lectures}, in: T. Dauxois, S. Ruffo, and L. F. Cugliandolo, Editors, {\it Long-Range Interacting Systems}, Lecture Notes of the Les Houches Summer School: vol. 90, August 2008 (Oxford University Press, Oxford, 2010).

\bibitem{G12PS} P. Gaspard, Phys. Scr. {\bf 86}, 058504 (2012).

\bibitem{G12JSM} P. Gaspard, J. Stat. Mech.: Th. Exp., P08021 (2012).

\bibitem{HPPG11} P.~I.~Hurtado, C.~P.~Espigares, J.~J.~del~Pozo, and P.~L.~Garrido, Proc. Natl. Acad. Sci. U.S.A. {\bf 108}, 7704 (2011).

\bibitem{HPPG14} P.~I.~Hurtado, C.~P.~Espigares, J.~J.~del~Pozo, and P.~L.~Garrido, J. Stat. Phys. {\bf 154}, 214 (2014).

\bibitem{HPPG15} C.~P.~Espigares, F.~Redig, and C.~Giardin{\'a}, J. Phys. A: Math. Theor. {\bf 48}, 35FT01 (2015).

\bibitem{LG14} D.~Lacoste and P.~Gaspard, Phys. Rev. Lett. {\bf 113}, 240602 (2014).

\bibitem{E85} R.~S.~Ellis, {\it Entropy, Large Deviations, and Statistical Mechanics} (Springer, New York, 1985).

\bibitem{E95} R.~S.~Ellis, Scand. Actuar. J. {\bf 1}, 97 (1995).

\bibitem{D07} B.~Derrida, J. Stat. Mech.: Th. Exp. P07023 (2007).

\bibitem{T09} H.~Touchette, Phys. Rep. {\bf 478}, 1 (2009).

\bibitem{P03} L.~Peliti, {\it Statistical Mechanics in a Nutshell} (Princeton University Press, Princeton and Oxford, 2003).

\bibitem{DL15} J. Guioth, and D. Lacoste, in preparation (2015).

\bibitem{H89} M. Hamermesh, {\it Group Theory and its Application to Physical Problems} (Dover, New York, 1989).

\bibitem{CDL91} C. Cohen-Tannoudji, B. Diu, and F. Laloe, {\it Quantum Mechanics}, Vol. I \& II (Wiley, New-York, 1991).

\bibitem{LL60} L. D. Landau and E. M. Lifshitz, {\it Electrodynamics of Continuous Media}, Course of Theoretical Physics, Vol. 8 (Pergamon Press, Oxford, 1960).

\bibitem{W65} B. Widom, J. Chem. Phys. {\bf 43}, 3898 (1965).

\bibitem{F67} M. E. Fisher, Rep. Prog. Phys. {\bf 30}, 615 (1967).

\bibitem{KGHHLPRSAK67} L.~P.~Kadanoff, W.~G\"otze, D.~Hamblen, R.~Hecht, E.~A.~S.~Lewis, V.~V.~Palciauskas, M.~Rayl, J.~Swift, D.~Aspnes, and J.~Kane, Rev. Mod. Phys. {\bf 39}, 395 (1967).

\bibitem{H87} K. Huang, {\it Statistical Mechanics}, 2nd edition (Wiley, New York, 1987).

\bibitem{LY52} T. D. Lee and C. N. Yang, Phys. Rev. {\bf 87}, 410 (1952).

\bibitem{CDR09} A. Campa, T. Dauxois, and S. Ruffo, Phys. Rep. {\bf 480}, 57 (2009). 
 
\bibitem{LL01} D.~Lacoste and T.~C.~Lubensky, Phys. Rev. E {\bf 64}, 041506 (2001).

\bibitem{Footnote}  This is the correct equation, unlike the one which appeared below Eq.~(15) of Ref.~\cite{LG14} 
without the term $-\beta f(\B)$.

\bibitem{BHL75} M. Blume, P. Heller, and N. A. Lurie, Phys. Rev. B {\bf 11}, 4483 (1975).

\bibitem{KT73} J.~M.~Kosterlitz and D.~J.~Thouless, J. Phys. C: Solid State Phys. {\bf 6}, 1181 (1973).

\bibitem{BHP98} S.~T.~Bramwell, P.~C.~W.~Holdsworth, and J.-F.~Pinton, Nature {\bf 396}, 552 (1998).

\bibitem{PHSB01} B.~Portelli, P.~C.~W.~Holdsworth, M.~Sellitto, and S.~T.~Bramwell, Phys. Rev. E {\bf 64}, 036111 (2001).

\bibitem{TC79} J. Tobochnik and G. V. Chester, Phys. Rev. E {\bf 20}, 3761 (1979).

\bibitem{M84} D. C. Mattis, Phys. Lett. A {\bf 104}, 357 (1984).

\bibitem{W82} F. Y. Wu, Rev. Mod. Phys. {\bf 54}, 235 (1982).

\bibitem{VHT04} R. Villavicencio-Sanchez, R. J. Harris, and H. Touchette, Eur. Phys. Lett. {\bf  105}, 30009 (2014).

\bibitem{dGP93} P.~G.~de~Gennes and J.~Prost, {\it The Physics of Liquid Crystals} (Oxford Science Publications, Oxford, 1993).

\bibitem{Pi81} S. A. Pikin, {\it Structural Transitions in Liquid Crystals} (Moskow: Nauka) (in russian) (1981).

\bibitem{AnZa82} N. Angelescu and V. A. Zagrebnov, J. Phys. A: Math. Gen. {\bf 15}, L639 (1982).

\bibitem{MS58} W.~Maier and Z.~Saupe, Zeitschrift Naturforsch. A {\bf 13}, 564 (1958).

\bibitem{BMA10} A.~K.~Bhattacharjee, G.~I.~Menon, and R.~Adhikari, J. Chem. Phys. {\bf 133}, 044112 (2010).

\bibitem{GECL69} Groupe d'{\'e}tude des cristaux liquides d'Orsay, J. Chem. Phys. {\bf 51}, 816 (1969).

\bibitem{AGCGJP08} D.~Andrieux, P.~Gaspard, S.~Ciliberto, N.~Garnier, S.~Joubaud, and A.~Petrosyan, J. Stat. Mech.: Th. Exp. P01002 (2008)

\bibitem{TKVBMDLB14} S.~Tusch, A.~Kundu, G.~Verley, T.~Blondel, V.~Miralles, D.~D{\'e}moulin, D.~Lacoste, and J.~Baudry, Phys. Rev. Lett. {\bf 112}, 180604 (2014).

\bibitem{JPCG08} S.~Joubaud, A.~Petrosyan, S.~Ciliberto, and N.~B.~Garnier, Phys. Rev. Lett. {\bf 100}, 180601 (2008).

\bibitem{W88} S. Wolfram, {\it Mathematica} (Addison-Wesley Publishing Company, Redwood City CA, 1988).

\bibitem{P65} C.~E.~Porter, Editor, {\it Statistical Theories of Spectra: Fluctuations} (Academic Press, New York, 1965).

\end{thebibliography}
\end{document}